\documentclass[article,authoryear,aaaaa,equation]{beg_32_ARXIV}    

\usepackage{graphicx}                           
    \DeclareGraphicsExtensions{.pdf,.jpg,.png}  
    \graphicspath{{chapter1/}{chapter2/}{chapter3/}{chapter4/}}

\usepackage{tikz} 
\usetikzlibrary{arrows, decorations.markings} 
\usetikzlibrary{shapes,matrix,chains,positioning, patterns,decorations.pathreplacing,arrows, calligraphy}
\usetikzlibrary{math}

\usepackage{amsmath, amsbsy}
\usepackage{amsfonts}
\usepackage{amsthm}
\usepackage{bbm}

\usepackage{pgfplots}    
\pgfplotsset{compat=newest} 
\usepgfplotslibrary{units} 

\usepackage{listings}                           
\usepackage{enumitem}                           
\usepackage{gensymb}			
\usepackage{subfiles}                           
\usepackage{subcaption}

\tikzset{cross/.style={cross out, draw=black, minimum size=2*(#1-\pgflinewidth), inner sep=0pt, outer sep=0pt},
cross/.default={6pt}}

\usepackage{tabularx}   
\usepackage{longtable} 
\usepackage{xltabular} 
\usepackage{booktabs}   
\usepackage[normalem]{ulem}
\usepackage{siunitx}   

\DeclareMathAlphabet\mathbfcal{OMS}{cmsy}{b}{n}

\usepackage{comment}

\usepackage{glossaries}
\newacronym{ad}{AD}{automatic differentiation}
\newacronym{adam}{ADAM}{adaptive moment estimation}
\newacronym{ai}{AI}{artificial intelligence}
\newacronym{ann}{ANN}{artificial neural network}
\newacronym{bc}{BC}{boundary condition}
\newacronym{bvp}{BVP}{boundary value problem}
\newacronym{cad}{CAD}{computer-aided design}
\newacronym{cpu}{CPU}{central processing unit}
\newacronym{dg}{DG}{Discontinuous Galerkin}
\newacronym{dgm}{DGM}{Deep Galerkin Method}
\newacronym{drm}{DRM}{Deep Ritz Method}
\newacronym{dof}{dof}{degrees of freedom}
\newacronym{em}{EM}{electromagnetic}
\newacronym{fc}{FC}{fully-connected}
\newacronym{fem}{FEM}{finite element method}
\newacronym{fnn}{FNN}{feedforward neural network}
\newacronym{gpu}{GPU}{graphics processing unit}
\newacronym{iga}{IGA}{isogeometric analysis}
\newacronym{lhs}{LHS}{Latin hypercube sampling}
\newacronym{mc}{MC}{Monte Carlo}
\newacronym{ml}{ML}{machine learning}
\newacronym{mse}{MSE}{mean squared error}
\newacronym{mlp}{MLP}{multilayer perceptron}
\newacronym{nurbs}{NURBS}{non-uniform rational B-splines}
\newacronym{ode}{ODE}{ordinary differential equation}
\newacronym{pde}{PDE}{partial differential equation}
\newacronym{pdf}{PDF}{probability density function}
\newacronym{pinn}{PINN}{physics-informed neural network}
\newacronym{pi}{PI}{physics-inspired}
\newacronym{pmsm}{PMSM}{permanent magnet synchronous machine}
\newacronym{qmc}{QMC}{quasi Monte Carlo}
\newacronym{sgd}{SGD}{stochastic gradient descent}

\usepackage{color}

\DeclareMathOperator*{\argmin}{argmin}

\usepackage[hang]{footmisc}
\fancypagestyle{plain}{%
  \fancyhf{}
  \fancyhead[R]{}
  \fancyfoot[R]{\small\bf\thepage }
  \fancyfoot[L]{}
  }
\renewcommand{\dmyy}{20}
\renewcommand{\myyear}{2021}
\renewcommand{\today}{}
\begin{document}

\volume{}
\title{A Neural Solver for Variational Problems on CAD Geometries with Application to Electric Machine Simulation}
\titlehead{Variational Neural Solver for CAD Geometries}
\authorhead{M. von Tresckow, S. Kurz, H. De Gersem, \& D. Loukrezis}
\corrauthor[1]{Moritz von Tresckow}
\author[2]{Stefan Kurz}
\author[1]{Herbert De Gersem}
\author[1]{Dimitrios Loukrezis}
\corremail{moritz.von\_tresckow@tu-darmstadt.de}
\corraddress{Technische Universit\"at Darmstadt, Institute for Accelerator Science and Electromagnetic Fields, Schlossgartenstr. 8, 64289 Darmstadt, Germany}
\address[1]{Technische Universit\"at Darmstadt, Institute for Accelerator Science and Electromagnetic Fields, Schlossgartenstr. 8, 64289 Darmstadt, Germany}
\address[2]{University of Jyv\"askyl\"a, Faculty of Information Technology, Seminaarinkatu 15, Jyv\"askyl\"a, Finland}



\abstract{
This work presents a deep learning-based framework for the solution of partial differential equations on complex computational domains described with computer-aided design tools.
To account for the underlying distribution of the training data caused by spline-based projections from the reference to the physical domain, a variational neural solver equipped with an importance sampling scheme is developed, such that the loss function based on the discretized energy functional obtained after the weak formulation is modified according to the sample distribution.
To tackle multi-patch domains possibly leading to solution discontinuities, the variational neural solver is additionally combined with a domain decomposition approach based on the Discontinuous Galerkin formulation.  
The proposed neural solver is verified on a toy problem and then applied to a real-world engineering test case, namely that of electric machine simulation. 
The numerical results show clearly that the neural solver produces physics-conforming solutions of significantly improved accuracy.
}


\maketitle

\section{Introduction}
\label{sec:intro} 
Recent years have seen a resurgence of numerical methods based on (deep) \glspl{ann} for solving ordinary (\acrshortpl{ode}) and \glspl{pde} \citep{budkina2016neural, chen2018neural, e2017deep, weinan2018deep, khodayi2020varnet, long2018pde, long2019pde, mall2016application, raissi2019physics, sirignano2018dgm, yadav2015introduction}.
These so-called \emph{neural solvers} revisit an idea with origins more than 20 years ago \citep{aarts2001neural, dissanayake1994neural, lagaris1998artificial, lagaris2000neural, lee1990neural, meade1994numerical} in the new light of the ongoing advances in \gls{ml} technologies, the availability of deep learning software \citep{abadi2016tensorflow, al2016theano, lu2021deepxde, haghighat2021sciann, paszke2017automatic}, and the capabilities of modern computing hardware \citep{jouppi2018motivation, lecun20191}.
Nevertheless, the use of neural solvers remains limited compared to standard numerical techniques such as the \gls{fem} \citep{hackbusch2017elliptic, strang1988analysis}. 
This can be attributed to a number of still unresolved issues, e.g., the lack of theoretical guarantees in terms of convergence and stability \citep{beck2019full, shin2020convergence}, or the typically heuristic and often cumbersome procedure of hyperparameter tuning \citep{wang2021understanding, kim2021dpm}.
However, there exist problem settings where neural solvers can be advantageous to standard numerical solvers, e.g. considering complex geometries \citep{berg2018unified} or high dimensions \citep{han2017solving, hutzenthaler2020proof}.

The core idea behind developing a neural solver is to recast the approximation of an \glsdisp{ode}{ODE} or \gls{pde} as an optimization problem, which is the natural setting of \gls{ann} training and thus \gls{ann}-based approximation. 
Therein, the main task is to design a suitable loss function, such that the \gls{ann} is trained to approximate the solution of the differential equation. 
Importantly, neural solvers approximate \glsdisp{ode}{ODE}/\gls{pde} solutions without having access to solution instants for a set of input data, as in the more common setting of supervised learning. 
Instead, the loss functions are designed such that the \gls{ann} is trained to conform to the physics of the problem described by the differential equation. 
Several approaches have been proposed in the literature for that purpose. 
In this work we focus on variational neural solvers for \glspl{pde} \citep{weinan2018deep, kharazmi2021hp, khodayi2020varnet}. In this case, the variational (weak) form of the \gls{pde} is first obtained and an \gls{ann} is used to approximate the trial functions.
Then, the energy functional obtained after the weak form is discretized according to a numerical quadrature rule, e.g. \gls{mc} or \gls{qmc} integration \citep{caflisch1998monte, lemieux2009monte}. 
The discretized energy functional constitutes the loss function to be minimized during model learning, which is typically accomplished with a gradient-based optimization algorithm such as \gls{sgd} \citep{bottou1991stochastic} or \gls{adam} \citep{kingma2014adam}.
Accordingly, the integration nodes or samples used to discretize the energy functional constitute the training data set. 
Other approaches construct loss functions based on the strong form of the \gls{pde} \citep{berg2018unified, raissi2019physics, sirignano2018dgm}. 
In this case it is often necessary to provide additional methods for the computation of higher-order derivatives, see e.g. \citep{berg2018unified, sirignano2018dgm}, as the standard tools of \gls{ad} \citep{baydin2015automatic} might become too cumbersome.
In either case, \glspl{bc} are incorporated by imposing hard or soft constraints on the minimization problem \citep{lagaris1998artificial, liao2019deep, mcfall2009artificial, raissi2019physics}.

This work is concerned with the application of variational neural solvers for the solution of \glspl{pde} on computational domains described with \gls{cad} tools, which are commonly employed in engineering design applications.
Typically in \gls{cad} geometry modeling, B-splines or \gls{nurbs} are used to define a projection map from a reference domain to the physical domain, which allows for an exact representation of the geometry \citep{braibant1984shape, hughes2005isogeometric}.
For complex geometries that cannot be represented by a single transformation of the reference domain in their entirety, a multi-patch approach is typically employed, which corresponds to a decomposition of the domain into non-overlapping subdomains (patches), each of which is represented through a dedicated projection map \citep{buffa2015approximation, langer2015multipatch}.
In this context, this work aims to address two main problems which arise considering \gls{ann}-based \gls{pde} approximations on \gls{cad} geometries.

The first problem tackled in this work occurs due to the fact that the training data on the physical geometry must be generated by first sampling the reference domain and then projecting the sampling points onto the physical domain. 
Then, the projected sampling points follow a distribution dependent on the projection map, which is different than the distribution used to sample the reference domain. 
This distribution must be taken into account during the discretization of the energy functional, equivalently, for the definition of the loss function.
There lies the first contribution of this work, which proposes an importance sampling scheme \citep{tokdar2010importance} for discretizing the energy functional, such that the loss function is chosen in accordance to the underlying probability distribution. 

The second problem addressed in this work concerns the case of decomposed (multi-patch) \gls{cad} geometries, which can possibly lead to sharp changes in the solution, e.g. considering adjacent subdomains with material discontinuities. 
In this case, it is crucial that the \gls{ann}-based approximation is able to resolve the subdomain interface conditions correctly.  
There lies the second contribution of this work, which proposes a domain decomposition approach along the lines of the \gls{dg} method \citep{brezzi2000discontinuous, cockburn2012discontinuous}, to be integrated within the variational neural solver.
In particular, one \gls{ann} is assigned per non-overlapping subdomain and the energy functional is given according to the \gls{dg} formulation. 
The corresponding \gls{dg}-based loss function integrates the contributions from all subdomains, along with additional terms that ensure the satisfaction of the \glspl{bc} and the interface conditions.

With respect to the latter contribution of this work, we note that recent works have also explored the combination of neural solvers and domain decomposition methods, albeit not in the context of \gls{cad} and without making use of the \gls{dg} method.
A shared idea is the use of multiple \glspl{ann}, i.e. one per subdomain. 
In \citep{li2020deep, li2019d3m} the computational domain is partitioned into overlapping subdomains, such that the \glspl{ann} are first trained separately and then updated depending on solution agreement on the intersections of the subdomains. 
Non-overlapping subdomains are used in \citep{jagtap2020conservative, jagtap2020extended}, where physics-based coupling factors are employed to enforce the interface conditions.
An exception to the use of multiple \glspl{ann} is the method presented in \citep{kharazmi2021hp}, which approximates the \gls{pde} with a single \gls{ann}, but making use of a separate set of test functions per subdomain.  
An $hp$-element-like approximation method based on partition of unity networks is suggested in \citep{lee2021partition, trask2021probabilistic}, however, this method concerns a purely data-driven, supervised learning context, where ground truth values are assumed to exist for the function to be approximated, equivalently, for the solution of the \gls{pde}.
Thus, the variational neural solver proposed in this paper is distinctly different than the ones presented in the aforementioned works.

The neural solver developed in this work is first verified on a toy problem from computational electromagnetics and then applied to a real-world engineering test case, namely, for the simulation of a \gls{pmsm} \citep{bhat2018modelling, bontinck2018isogeometric, bontinck2018robust, ion2018robust, merkel2021shape}. 
The real-world test case features all problems that this work aims to address, in particular, a complicated, multi-patch \gls{cad} geometry described by means of \gls{nurbs}, upon which materials with different \gls{em} properties coexist. 
Aside from the aforementioned methodological contributions, this is the first time a neural solver is employed to simulate an electric machine, at least to the authors' knowledge.

The rest of this paper is organized as follows.
Section \ref{sec:neural-solver} details the development of a variational neural solver, using a model problem. Implementation specifics that hold throughout this work are also mentioned in this section.
In section \ref{sec:importance-sampling}, the basic tools of \gls{cad} geometry representation are presented. In the same section, an importance sampling scheme is derived, which is used to appropriately discretize the energy functional given the projection-dependent sample distribution and accordingly choose the loss function to be minimized during \gls{ann} training. 
Section \ref{sec:dg} considers the case of decomposed \gls{cad} domains and further modifies the loss function based on the \gls{dg} formulation.
Numerical experiments which showcase the advantages of the proposed neural solver are available in section \ref{sec:num-results}.
A discussion on the findings of the present work as well as a research outlook are available in section \ref{sec:conclusion}.

\section{Variational neural solver}
\label{sec:neural-solver}

Variational neural solvers are \gls{ann}-based approximation methods, which are particularly suited for energy functional minimization problems arising from variational \gls{pde} formulations \citep{weinan2018deep, kharazmi2021hp, khodayi2020varnet}. 
Postponing the discussion regarding \glspl{bc} until section \ref{subsec:bcs}, we consider the elliptic \gls{pde}
\begin{equation}
	\label{eq:poisson_no_bcs}
	-\Delta u\left(\mathbf{x}\right) = f\left(\mathbf{x}\right), \:\:\:\mathbf{x} \in \Omega,
\end{equation}
where $\Omega \subset \mathbb{R}^D$ denotes the computational domain and $\mathbf{x}$ is the spatial coordinate vector.
Applying the standard tools for the variational formulation \citep{evans2012numerical, hackbusch2017elliptic}, the solution to \eqref{eq:poisson_no_bcs} is given by minimizing the energy functional
\begin{equation}
	\label{eq:energy_functional}
	I\left(u\left(\mathbf{x}\right)\right) = \int_{\Omega}  \left(\frac{1}{2} \left|\nabla u\left(\mathbf{x}\right)\right|^2 - f\left(\mathbf{x}\right) u\left(\mathbf{x}\right)\right) \text{d}\mathbf{x},
\end{equation}
where $u$ are trial functions living in the Sobolev space $H^1(\Omega)$, and  $f$ is a forcing function. 
Then, the solution of the \gls{pde} is given as $u^*(\mathbf{x}) = \argmin_{u \in H^1\left(\Omega\right)} I\left(u\left(\mathbf{x}\right)\right)$.  
In the following, we drop the dependency on the spatial coordinate vector $\mathbf{x}$ wherever possible in order to simplify the notation.
The main idea behind variational neural solvers is to train an \gls{ann} to minimize the energy functional \eqref{eq:energy_functional}, and use this \gls{ann} as an approximation to the solution of the \gls{pde}, as presented in more detail in sections \ref{subsec:ann-approx} -- \ref{subsec:bcs}.

\subsection{ANN-based approximation of trial functions}
\label{subsec:ann-approx}
The trial functions $u$ in \eqref{eq:energy_functional} are approximated by an \gls{ann}, most commonly a \gls{mlp}.
To define the \gls{mlp}, we must first introduce the concepts of \emph{neuron} and \emph{layer}.
A neuron is the smallest unit of an \gls{ann} and is mathematically described by the function
\begin{equation}
	\label{eq:lin_class}
	\mathcal{N}\left(\mathbf{x}\right)=\sigma\left(\mathbf{x}^\top \mathbf{w}  + b\right),
\end{equation}
where $\mathbf{x} \in \mathbb{R}^D$ is the input vector, $\mathbf{w}\in \mathbb{R}^D$ a weight vector, $b \in \mathbb{R}$ a bias term, and $\sigma: \mathbb{R} \rightarrow \mathbb{R}$ a nonlinear function known as the \emph{activation}. 
A set of neurons which receive the same input form a layer, the output of which is given by
\begin{equation}
	\mathcal{L}\left(\mathbf{x}\right) = \sigma \left( \mathbf{x}^\top \mathbf{W} + \mathbf{b} \right), 
\end{equation}
where $\mathbf{W} \in \mathbb{R}^{D \times N}$ is a matrix concatenating the weights of the individual neurons, $\mathbf{b} \in \mathbb{R}^N$ is the corresponding bias vector, and the activation function $\sigma$ is now applied element-wise for each neuron of the layer.
An \gls{mlp} is a specific type of \gls{ann} formed by a sequence of layers and thus described via the layer composition formula
\begin{equation}
	\label{eq:recur_ANN}
	\mathcal{MLP}\left(\mathbf{x}\right) = \mathcal{L}^{(L)} \left( \mathcal{L}^{(L-1)} \left( \cdots  \left( \mathcal{L}^{(1)} \left(\mathbf{x}\right) \right) \right) \right),
\end{equation}
where $L$ denotes the number of layers in sequence.
The layers $\mathcal{L}^{(l)}$, $l=1, \dots L-1$, are called the hidden layers and their number defines the depth of the \gls{ann}.
The weights and biases $\boldsymbol{\theta} = \left\{\mathbf{W}^{(l)},\mathbf{b}^{(l)}\right\}_{l=1}^L$ of the \gls{ann} are called the trainable parameters.
We denote the approximation of the trial functions by the \gls{ann} with $u \approx u_{\boldsymbol{\theta}}$.

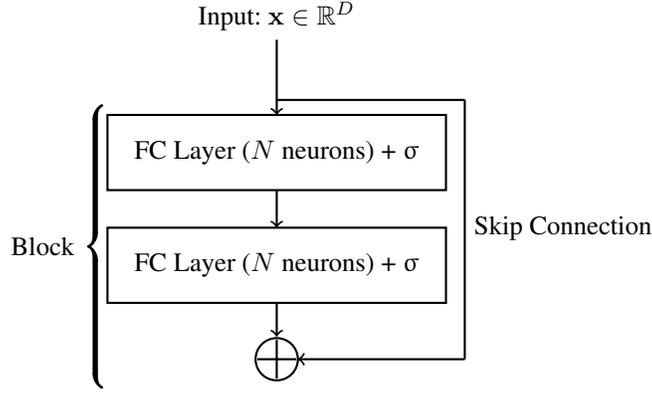
\begin{figure}[t!]
	\begin{center}
		\begin{tikzpicture}
		\draw (0,0) node[anchor=south]{Input: $\mathbf{x} \in \mathbb{R}^D$};
		\foreach \shift in {0}{
			\draw[thick, ->] (0,0-\shift) -- (0,-1-\shift);
			\draw[thick](0,-0.8-\shift) -- (2.5, -0.8-\shift);
			\draw[thick, ->](2.5, -4.25-\shift) --(8pt,-4.25-\shift);
			\draw[thick](2.5, -0.8-\shift) --(2.5,-4.25-\shift);
			\draw (2.5, -2.5-\shift) node[anchor=west]{Skip Connection};
			\draw[thick] (-2.25,-2-\shift) rectangle (2.25,-1-\shift);
			\draw (0,-1.5-\shift) node[anchor=center]{\gls{fc} Layer ($N$ neurons) + $\sigma$};
			\draw[thick, ->] (0,-2-\shift)--(0,-2.5-\shift);
			\draw[thick] (-2.25,-3.5-\shift) rectangle (2.25,-2.5-\shift);
			\draw (0,-3-\shift) node[anchor=center]{\gls{fc} Layer ($N$ neurons) + $\sigma$};
			\draw[thick, ->] (0,-3.5-\shift)--(0,-3.95-\shift);
			\draw[thick] (0,-4.25-\shift) node[cross, rotate=45]{};
			\draw[thick] (0,-4.25-\shift) circle (8pt);
		};
		\draw (-2.2,-0.75) node[anchor=east](point1){};
		\draw (-2.2,-4.75) node[anchor=east](point2){};
		\draw[very thick, decoration={calligraphic brace, mirror, amplitude=5pt},decorate, line width=2pt]
		(point1) -- node[left=6pt] {Block} (point2);
		\end{tikzpicture}
		\caption{A single block of the \gls{ann} structure employed in this work. We denote with $\sigma$ the activation function, $N$ is the number of neurons in each \glsfirst{fc} layer and $\mathbf{x}$ is the input vector. The output of a block becomes the input of the next block in sequence.}
		\label{fig:ritz_structure}
	\end{center}
\end{figure}

The success of neural solvers, variational or not, depends crucially on \gls{ann} architecture, that is, number of layers, layer connection, neurons per layer, and activation functions. 
The architecture defines the expressive capabilities of the \gls{ann} and determines which functions it is able to approximate \citep{cybenko1989approximation, hornik1991approximation}. 
This work adopts the network structure proposed in \citep{weinan2018deep}, such that the \glspl{ann} consist of block sequences, with each block consisting of two \gls{fc} layers and a skip connection, see figure \ref{fig:ritz_structure}.
The skip connection is used to prevent the so-called vanishing gradient problem \citep{hochreiter2001gradient, pascanu2013difficulty}. 
The number of neurons per \gls{fc} layer as well as the number of blocks are chosen according to the specific requirements of the application at hand.
Additionally, we employ adaptive activation functions of the form $\sigma_l (x) = \tanh(a_l x)$, $l=1, \dots, L-1$, where $a_l \in \mathbb{R}$ are layer-based trainable parameters to be optimized along with the weights and biases. 
The scalable parameters $a_l$ affect the slope of the activation function, which is an important factor regarding the approximation capability of an \gls{ann}, and have been found to improve \gls{ann} performance in terms of approximation accuracy, robustness, and convergence \citep{jagtap2020adaptive, jagtap2020locally}.

\subsection{Energy functional discretization and variational loss function} 
\label{en-fun-discrete}
The loss function to be minimized during \gls{ann} training is obtained by discretizing the energy functional \eqref{eq:energy_functional} using numerical integration. 
Accordingly, the quadrature nodes constitute the training data.
Most commonly, the energy functional is discretized using pseudo-random samples $\left\{\mathbf{x}_m\right\}_{m=1}^M \subset \Omega$, in the spirit of \gls{mc} or \gls{qmc} integration \citep{e2017deep, han2017solving, sirignano2018dgm}, however, Gauss quadrature has also been employed \citep{kharazmi2021hp}.
The discretization method of choice in this work is \gls{qmc} sampling based on Sobol sequences \citep{caflisch1998monte, lemieux2009monte}, hence, the discretized energy functional under the uniform sampling assumption reads
\begin{equation}
	\label{eq:energy_functional_discrete}
	I(u) \approx  I\left(u_{\boldsymbol{\theta}}\right) \approx \widetilde{I}\left(u_{\boldsymbol{\theta}}\right) = \frac{1}{M}\sum_{m=1}^M \left( \frac{1}{2} \left|\nabla u_{\boldsymbol{\theta}}\left(\mathbf{x}_m\right)\right|^2 - f\left(\mathbf{x}_m\right) u_{\boldsymbol{\theta}}\left(\mathbf{x}_m\right)\right).
\end{equation}
Note that the integral estimator \eqref{eq:energy_functional_discrete} will be later corrected according to the importance sampling scheme presented in section \ref{subsec:importance}, in order to account for sample distributions induced by \gls{cad} tools.

\subsection{ANN training and PDE solution}
\label{subsec:ann-training}
A gradient-based optimization algorithm is employed to minimize the discretized energy functional \eqref{eq:energy_functional_discrete}.  
In this work, we use the \gls{adam} algorithm \citep{kingma2014adam}, which updates iteratively the parameter vector $\boldsymbol{\theta}$ as 
\begin{equation}
	\label{eq:adam_update}
	\boldsymbol{\theta}_{k+1} = \boldsymbol{\theta}_{k} - \eta \frac{\mathbf{\hat{m}}_{k+1}}{\sqrt{\mathbf{\hat{v}}_{k+1}+\epsilon}},
\end{equation}
where $\eta \in \mathbb{R}$ is the learning rate, $\mathbf{\hat{m}}_{k+1}$ the bias-corrected first moment estimate, $\mathbf{\hat{v}}_{k+1}$ the bias-corrected second moment estimate, and $\epsilon>0$ a small scalar that prevents division by zero. 
All operations in \eqref{eq:adam_update} are performed element-wise.
The initial values of the \gls{ann}'s parameter vector $\boldsymbol{\theta}$ are obtained using the so-called Xavier initialization \citep{glorot2010understanding}.
Note that the computation of the gradients with respect to either the spatial coordinates or the trainable parameters of the \gls{ann} can be performed by means of \gls{ad}, typically using the backpropagation algorithm \citep{baydin2015automatic, goodfellow2016deep}.
Further note that no mini-batching is employed in this work, i.e. the full training data set is used in every iteration of the \gls{adam} algorithm. 
Hence, we refer to a single iteration $k$ in \eqref{eq:adam_update} as an ``epoch'', which is the standard term used in \gls{ml} literature to indicate the number of passes of the entire training dataset completed by the training algorithm.
The approximate minimizer of \eqref{eq:energy_functional} is obtained by the \gls{ann} $u_{\boldsymbol{\theta^*}}$ which fulfills
\begin{equation}
	\widetilde{I}\left(u_{\boldsymbol{\theta^*}}\right) \leq \widetilde{I}\left(u_{\boldsymbol{\theta}}\right), \:\:\forall \theta \subset \mathbb{R}^{\left|\boldsymbol{\theta}\right|},
\end{equation}
where $\left|\boldsymbol{\theta}\right|$ denotes the dimension of the parameter vector.
The overall optimization problem can be summarized to:
\begin{equation}
	\label{eq:opti}
	\text{Find} \:\: \boldsymbol{\theta}^* \in  \underset{\boldsymbol{\theta} \subset \mathbb{R}^{\left|\boldsymbol{\theta}\right|}} {\operatorname{argmin}} \, \widetilde{I}\left(u_{\boldsymbol{\theta}}\right).
\end{equation}

\subsection{Inclusion of Boundary Conditions}
\label{subsec:bcs}
We now complement the \gls{pde} \eqref{eq:poisson_no_bcs} with Neumann and Dirichlet \glspl{bc}, thus obtaining the \gls{bvp}  
\begin{subequations}
	\label{eq:poisson_dir_neu}
	\begin{align}
		-\Delta u &= f, &&\mathbf{x} \in \Omega, \\ 
		u &= g_{\text{D}}, &&\mathbf{x} \in \Gamma_{\text{D}},\\ 
		\mathbf{n}_{\Gamma_{\text{N}}} \cdot \nabla u &= g_\text{N}, &&\mathbf{x} \in \Gamma_{\text{N}},
	\end{align}
\end{subequations}
where we assume that the boundary $\partial \Omega$ of the computational domain can be separated into a Neumann boundary $\Gamma_{\text{N}}$ and a Dirichlet boundary $\Gamma_{\text{D}}$, such that $\partial \Omega = \Gamma_{\text{D}} \cup \Gamma_{\text{N}}$ with $\Gamma_{\text{D}} \cap \Gamma_{\text{N}} = \emptyset$, and $\mathbf{n}_{\Gamma_{\text{N}}}$ denotes the outer normal vector on $\Gamma_{\text{N}}$.

Neumann \glspl{bc} are naturally present in the variational formulation, simply modifying the energy functional as
\begin{equation}
	\label{eq:energy_functional_neu}
	I\left(u\right) = \int_{\Omega} \left(\frac{1}{2} \left|\nabla u\right|^2 - f u\right)\mathrm{d}\mathbf{x} - \int_{\Gamma_\text{N}} u g_\text{N} \mathrm{d}\mathbf{x}.
\end{equation}
Thus, Neumann \glspl{bc} do not have to be explicitly accounted for, other than using separate sets of integration samples within $\Omega$ and on $\Gamma_{\text{N}}$ in the discretization of \eqref{eq:energy_functional_neu}, such that
\begin{equation}
	\label{eq:energy_functional_discrete_neu}
	I\left(u_{\boldsymbol{\theta}}\right) \approx \frac{1}{M}\sum_{m=1}^M  \left( \frac{1}{2} \left|\nabla u_{\boldsymbol{\theta}}\left(\mathbf{x}_m\right)\right|^2 - f\left(\mathbf{x}_m\right) u_{\boldsymbol{\theta}}\left(\mathbf{x}_m\right)\right) - \frac{1}{J}\sum_{j=1}^J  u_{\boldsymbol{\theta}}\left(\mathbf{x}_j\right) g_{\text{N}}\left(\mathbf{x}_j\right),
\end{equation}  
where $\left\{\mathbf{x}_m\right\}_{m=1}^M \subset \Omega$ and $\left\{\mathbf{x}_j\right\}_{j=1}^J \subset \Gamma_{\text{N}}$.

On the contrary, Dirichlet \glspl{bc} must be handled separately, as they impose constraints to the space of trial functions $u$.
In this work we use the option of soft constraints for including Dirichlet \glspl{bc}, in which case a penalty term is added to the energy functional \eqref{eq:energy_functional_neu}, forcing the \gls{ann} to learn the Dirichlet \glspl{bc} along with the \gls{pde} and the Neumann \glspl{bc} \citep{raissi2019physics, liao2019deep}.
The modified energy functional reads
\begin{align}
	\label{eq:energy_functional_neu_dir}
	I\left(u\right) = \int_{\Omega} \left(\frac{1}{2} \left|\nabla u\right|^2 - f u\right)\mathrm{d}\mathbf{x} 
	- \int_{\Gamma_\text{N}} u g_\text{N} \mathrm{d}\mathbf{x}
	+ \beta_{\text{D}} \int_{\Gamma_\text{D}}\left(u - g_\text{D}\right)^2\mathrm{d}\mathbf{x},
\end{align}
where $\beta_{\text{D}}$ is a scalar penalty factor.
Similar to the Neumann case, \eqref{eq:energy_functional_neu_dir} must be discretized using different integration samples in $\Omega$, on $\Gamma_{\text{N}}$, and on $\Gamma_{\text{D}}$, such that 
\begin{align}
	\label{eq:energy_functional_discrete_neu_dir}
	I\left(u_{\boldsymbol{\theta}}\right) \approx \frac{1}{M}\sum_{m=1}^M \left( \frac{1}{2} \left|\nabla u_{\boldsymbol{\theta}}\left(\mathbf{x}_m\right)\right|^2 - f\left(\mathbf{x}_m\right) u_{\boldsymbol{\theta}}\left(\mathbf{x}_m\right)\right) 
	&- \frac{1}{J_1}\sum_{j_1=1}^{J_1}  u_{\boldsymbol{\theta}}\left(\mathbf{x}_{j_1}\right) g_{\text{N}}\left(\mathbf{x}_{j_1}\right) \nonumber \\
	&+\beta_{\text{D}} \frac{1}{J_2}\sum_{j_2=1}^{J_2}  \left(u_{\boldsymbol{\theta}}\left(\mathbf{x}_{j_2}\right) - g_{\text{D}}\left(\mathbf{x}_{j_2}\right)\right)^2,
\end{align}  
where $\left\{\mathbf{x}_m\right\}_{m=1}^M \subset \Omega$, $\left\{\mathbf{x}_{j_1}\right\}_{j_1=1}^{J_1} \subset \Gamma_{\text{N}}$, and $\left\{\mathbf{x}_{j_2}\right\}_{j_2=1}^{J_2} \subset \Gamma_{\text{D}}$.

\section{CAD geometry representation and importance sampling}
\label{sec:importance-sampling}

Geometry representation in \gls{cad} is typically based on free-form curves, such that the computational domain is defined through a projection map from a reference domain to the physical domain.
Central to this procedure are B-splines and \glspl{nurbs}, presented in more detail in section \ref{subsec:splines}. 
Exactly due to this projection map, a standard \gls{mc} or \gls{qmc} discretization of the energy functional, e.g. as in the integral estimator \eqref{eq:energy_functional_discrete}, is not sufficient, due to the fact that the map induces a sample distribution on the physical domain which is different than the sample distribution on the reference domain. 
To alleviate this issue, we propose to discretize the energy functional using an importance sampling scheme, such that the loss function takes into account the sample distribution, as shown in section \ref{subsec:importance}.

\subsection{B-splines and NURBS}
\label{subsec:splines}
Assuming a \emph{knot vector} $\Xi = \left\{\xi_1, \dots, \xi_{n+p+1}\right\} \subset \left[0,1\right]$ such that $\xi_1 \leq \xi_2 \leq \cdots \leq \xi_{n+p+1}$, a basis of $n$ piece-wise polynomial functions of degree $p$ can be defined via the Cox-de-Boor recursion formula \citep{piegl1996nurbs}
\begin{equation}
	\label{eq:cox-de-boor}
	B_{i,p}(\xi)=\frac{\xi-\xi_{i}}{\xi_{i+p}-\xi_{i}}B_{i, p-1}(\xi) + \frac{\xi_{i+p+1} - \xi}{\xi_{i+p+1} - \xi_{i+1}}B_{i+1, p-1}(\xi), \quad i=1, \dots, n.
\end{equation}
The functions defined by formula \eqref{eq:cox-de-boor} are called \emph{B-splines}.
For $p=0$, a B-spline is defined as
\begin{equation}
	B_{i,0}(\xi)=
	\begin{cases}
		1, \quad \text{ if } \xi_1 \leq \xi < \xi_{i+1},\\
		0, \quad \text{otherwise.}	
	\end{cases}
\end{equation}
\Gls{nurbs} can be considered as a generalization of B-splines and are defined through the relation
\begin{equation}
	N_{i,p}=\frac{w_iB_{i,p}(\xi)}{\sum_{j=1}^n w_j B_{j,p}(\xi)},	
\end{equation}
where $w_i$ are positive weight factors. 
Using $n$ \gls{nurbs} as basis functions along with $n$ control points $\{\mathbf{P}_i \}_{i=1}^{n} \subset \Omega$, a \gls{nurbs} curve can be defined as
\begin{equation}
	\mathbf{C}(\xi)=\sum_{i=1}^n\mathbf{P}_i N_{i,p}(\xi).
\end{equation}
\gls{nurbs} surfaces can be created by tensor products of \gls{nurbs} curves.
Using these constructions, it is possible to define a smooth map
\begin{equation}
	\label{eq:cad-map}
	F:\left[0,1\right]^{d} \rightarrow \overline{\Omega} \subset \mathbb{R}^d, \quad d\in\left\{1,2,3\right\},	
\end{equation}
which parametrizes the physical geometry $\overline{\Omega} = \Omega \cup \partial \Omega$.
If $\overline{\Omega}$ is not a self-penetrating domain, $F$ is a bijective map \citep{kleiss2015two}.  
A crucial advantage is that the map $F$ can be easily obtained from any CAD software.

\subsection{Importance sampling-based energy functional discretization}
\label{subsec:importance}
We consider again the model problem \eqref{eq:poisson_no_bcs}, where we now assume that the (open-boundary) physical domain $\Omega$ is described via a map $F:\left[0,1\right]^2 \rightarrow \Omega$, as in section \ref{subsec:splines}.
The generation of samples on the physical domain proceeds in two steps. 
First, uniformly distributed sample points are generated on the unit square, such that $\left\{\mathbf{y}_{m}\right\}_{m=1}^{M} \subset \left[0,1\right]^2$. 
Then, the map $F$ is used to project these sample points on the physical geometry, i.e. $F\left(\left\{\mathbf{y}_{m}\right\}_{m=1}^{M}\right) = \left\{\mathbf{x}_{m}\right\}_{m=1}^{M} \subset \Omega$.

On the physical domain $\Omega$, the sample points are not uniformly distributed, but instead have an underlying distribution which is dependent on the map $F$.
We denote the corresponding \gls{pdf} with $p\left(\mathbf{x}\right)$.
To derive an explicit expression for $p\left(\mathbf{x}\right)$, we apply the transformation theorem for \glspl{pdf} under diffeomorphisms \citep{klenke2006wahrscheinlichkeitstheorie}. 
The theorem states that, for a random vector $\mathbf{y}=\left(y_1,...,y_n \right)$ with \gls{pdf} $p_{\mathbf{y}}\left(\mathbf{y}\right)$ and a diffeomorphism $f:\mathbb{R}^n \rightarrow \mathbb{R}^n$, the \gls{pdf} $p_{\mathbf{x}}$ of $\mathbf{x} = f\left(\mathbf{y}\right)$ is given by
\begin{equation}
	p_{\mathbf{x}}\left(\mathbf{x}\right) = p_{\mathbf{y}}\left(f^{-1}\left(\mathbf{x}\right)\right) \underbrace{\left|\frac{\partial \left(y_1,...,y_n\right)}{\partial \left(x_1,...,x_n\right)}\right|}_{:=\Phi_{f^{-1}}},
\end{equation}
where $\Phi_{f^{-1}}$ is the Jacobian determinant of $f^{-1}$.
The map $F$ is bijective and smooth, therefore, it constitutes a diffeomorphism, is invertible, and the Jacobian of the inverse exists.
The bijectivity property holds if the considered domain is not self-penetrating \citep{kleiss2015two}.
Since $F$ is bijective, we also know that $F^{-1}(\mathbf{x}_{m})\in \left[0,1\right]^2$, $\forall \mathbf{x}_{m}$, $m=1,\dots,M$.
Thus, the \gls{pdf} $p\left(\mathbf{x}\right)$ is given as
\begin{equation}
	p\left(\mathbf{x}\right) = \underbrace{p_{\mathbf{y}}\left(F^{-1}\left(\mathbf{x}\right)\right)}_{=1} \left|\frac{\partial \left(y_1,...,y_n\right)}{\partial \left(x_1,...,x_n\right)}\right|=\left|\frac{\partial \left(y_1,...,y_n\right)}{\partial \left(x_1,...,x_n\right)}\right|=\Phi_{F^{-1}}\left( \mathbf{y} \right).
\end{equation}
An importance sampling-based integral estimator can then be derived as
\begin{align}
	\label{eq:is_mc_est}
	\int_{\Omega} u\left(\mathbf{x}\right) \mathrm{d}\mathbf{x}
	\approx  \frac{1}{M}\sum_{m=1}^{M} \frac{u\left(\mathbf{x}_{m}\right)}{p\left(\mathbf{x}_{m}\right)} = \frac{1}{M} \sum_{m=1}^{M} u\left(\mathbf{x}_{m}\right) \Phi_{F}\left(\mathbf{x}_{m}\right),
\end{align}
where the relation $\Phi_{F^{-1}}=(\Phi_{F})^{-1}$ is used.
Accordingly, the discretized energy functional \eqref{eq:energy_functional_discrete} is modified to
\begin{equation}
	\label{eq:energy_functional_discrete_is}
	\widetilde{I}\left(u_{\boldsymbol{\theta}}\right) = \frac{1}{M}\sum_{m=1}^M \Phi_{F}\left(\mathbf{x}_{m}\right) \left( \frac{1}{2} \left|\nabla u_{\boldsymbol{\theta}}\left(\mathbf{x}_m\right) \right|^2 - f\left(\mathbf{x}_m\right) u_{\boldsymbol{\theta}}\left(\mathbf{x}_m\right) \right).
\end{equation}

The discretization of domain boundaries, e.g. considering the \gls{bvp} \eqref{eq:poisson_dir_neu}, proceeds analogously, hence, we omit the explicit form of the discretized energy functional including \glspl{bc}.
Effectively, using the importance sampling scheme for discretizing the energy functional, the terms of the loss function are weighted according to the underlying sample distribution induced by the projection map $F$.

\section{Domain decomposition and Discontinuous Galerkin formulation}
\label{sec:dg}

\subsection{Multi-patch CAD geometries and domain decomposition}
\label{subsec:multi-patch}
In most practical applications, the geometry cannot be parametrized using a single projection map from the reference domain to the physical domain.
In such cases, a multi-patch parameterization is employed, which essentially decomposes the computational domain $\Omega$ into non-overlapping subdomains $D_k$, $k=1,\dots,K$, such that $\Omega = \bigcup_{k=1}^K D_k$. 
We denote the boundary of each subdomain with $\partial D_k$. 
The interface between two adjacent subdomains $D_k$ and $D_l$ is denoted as $\Gamma_{k,l} = \Gamma_{l,k} = \partial D_k \cap \partial D_l$.
Assuming that the boundary $\partial \Omega$ consists of a Neumann boundary $\Gamma_{\text{N}} \subset \partial \Omega$ and a Dirichlet boundary $\Gamma_{\text{D}} \subset \partial \Omega$, if part of $\partial D_k$ happens to be on $\Gamma_{\text{N}}$ or on $\Gamma_{\text{D}} $, we denote this part as $\Gamma_{k, \text{N}} = \Gamma_{k} \cap \Gamma_{\text{N}}$ or $\Gamma_{k, \text{D}} = \Gamma_{k} \cap \Gamma_{\text{D}}$, respectively.

For such cases, instead of using a single \gls{ann} to approximate the \gls{pde} solution over the whole computational domain $\Omega$, we propose to train one \gls{ann} per subdomain $D_k$ and then combine the individual subdomain solutions in order to get the solution over the whole domain. 
However, allowing complete independence between the subdomain-specific \glspl{ann} could possibly result in nonphysical solutions, in particular ones that disregard interface conditions to be satisfied on $\Gamma_{k,l}$.
For that reason, it is necessary to modify the energy functional in order to account for the interface conditions as well.
To that end, methods using physics-based coupling factors and penalty terms have been suggested in the literature \citep{jagtap2020conservative, jagtap2020extended}.
In this work, we propose an alternative approach based on the \gls{dg} method, which is presented in section \ref{subsec:dg-loss}.

\subsection{Energy functional based on the Discontinuous Galerkin formulation}
\label{subsec:dg-loss}

Focusing on the model \gls{bvp} \eqref{eq:poisson_dir_neu} and assuming a decomposition of the domain $\Omega$ as presented in section \ref{subsec:multi-patch}, the energy functional is modified along the lines of a \gls{dg} formulation, such that both \glspl{bc} and interface conditions are described with average and jump operators \citep{brezzi2000discontinuous, cockburn2012discontinuous}.
The average operator on $u$ and $\nabla u$ is defined as
\begin{subequations}
	\label{eq:avg-operator}
	\begin{align}
		\left\{u\right\}_{\Gamma} &= \frac{1}{2}\left(u_k + u_l\right), &&\Gamma \equiv \Gamma_{k,l},\\
		\left\{u\right\}_{\Gamma} &= g_{\text{D}}, &&\Gamma \equiv \Gamma_{k, \text{D}}, \\
		\left\{u\right\}_{\Gamma} &= u_k, &&\Gamma \equiv \Gamma_{k, \text{N}}, \\
		\left\{\nabla u\right\}_{\Gamma} &= \frac{1}{2}\left(\nabla u_k + \nabla u_l\right), &&\Gamma \equiv \Gamma_{k,l},\\
		\left\{\nabla u\right\}_{\Gamma} &= \nabla u_k, &&\Gamma \equiv \Gamma_{k,\text{D}} \\
		\left\{\nabla u\right\}_\Gamma \cdot \mathbf{n}_{\Gamma} &= g_{\text{N}}, &&\Gamma \equiv \Gamma_{k,\text{N}},
	\end{align}
\end{subequations}
where $\mathbf{n}_{\Gamma}$ denotes the outer unit normal vector of $\Gamma$. 
The jump operator is defined as
\begin{align}
	\label{eq:jump-operator}
	\left[u\right]_{\Gamma} =
	\begin{cases}
		\left(u_k - u_l\right) \cdot \mathbf{n}_{\Gamma}, &\Gamma \equiv \Gamma_{k,l}, \\
		0, &\Gamma \equiv \Gamma_{k, \text{N}}, \\
		u_{k} \cdot \mathbf{n}_\Gamma &\Gamma \equiv \Gamma_{k, \text{D}}.
	\end{cases}	
\end{align}
Using the average and jump operators, the energy functional is modified to 
\begin{align}
	\label{eq:energy_functional_DG}
	I(u) &= \sum_{k=1}^K \int_{D_k} \left(\frac{1}{2} \left|\nabla u_k\right|^2 - fu_k\right) \mathrm{d}\mathbf{x}
	- \sum_{k=1}^K \int_{\Gamma_{k,\text{N}}} u_k g_{\text{N}} \mathrm{d}\mathbf{x} \nonumber \\
	&- \sum_{k=1}^K \sum_{l>k}^K \int_{\Gamma_{k,l}} \left\{\nabla u \right\}_{\Gamma_{k,l}}  \left[u\right]_{\Gamma_{k,l}} \mathrm{d}\mathbf{x} 
	+ \sum_{k=1}^K \sum_{l>k}^K \frac{\beta_{k,l}}{2} \int_{\Gamma_{k,l}} \left[u\right]_{\Gamma_{k,l}}^2 \mathrm{d}\mathbf{x} \nonumber \\
	&- \sum_{k=1}^K \int_{\Gamma_{k,\text{D}}} \left\{\nabla u \right\}_{\Gamma_{k, \text{D}}} \left[u\right]_{\Gamma_{k,\text{D}}} \mathrm{d}\mathbf{x} 
	+ \sum_{k=1}^K \frac{\beta_{k,\text{D}}}{2} \int_{\Gamma_{k,\text{D}}} \left[u - g_{\text{D}}\right]_{\Gamma_{k,\text{D}}}^2 \mathrm{d}\mathbf{x}, 
\end{align}
the discretization of which yields the loss function to be minimized during model training.
For the discretization of the \gls{dg}-based energy functional \eqref{eq:energy_functional_DG}, we use the importance sampling scheme derived in section \ref{subsec:importance}, which shall now be employed per subdomain $D_k$, $k=1,\dots,K$. 
Accordingly, subdomain-specific projection maps $F_k: \left[0,1\right]^2 \rightarrow D_k$ and \glspl{pdf} $p_k(\mathbf{x})$ are used.

\section{Numerical results}
\label{sec:num-results}
In the following, the variational neural solver developed in this work is first verified on a toy \gls{em} field problem and then applied for the purpose of electric machine simulation. 
In both test cases, the \gls{dg}-based variational neural solver is compared against a neural solver employing a single \gls{ann}, i.e. without any domain decomposition considerations, as well as against a domain decomposition approach which also utilizes one \gls{ann} per subdomain along with physics-inspired coupling factors to accommodate the interface conditions, similar to \citep{jagtap2020conservative, jagtap2020extended}.

In the latter approach, the coupling factors are implemented as additional penalty terms to the energy functional. Revisiting the model \gls{bvp} \eqref{eq:poisson_dir_neu} under the domain decomposition assumptions given in section \ref{subsec:multi-patch}, the modified energy functional reads
\begin{align}
	\label{eq:energy_functional_physics_interface}
	I(u) &= \sum_{k=1}^K \int_{D_k} \left(\frac{1}{2} \left|\nabla u_k\right|^2 - fu\right) \mathrm{d}\mathbf{x}
	- \sum_{k=1}^K \int_{\Gamma_{k,\text{N}}} u_k g_{\text{N}} \mathrm{d}\mathbf{x} \nonumber \\
	&+ \sum_{k=1}^K \beta_{k,\text{D}} \int_{\Gamma_{k, \text{D}}} \left(u_k - g_{\text{D}} \right)^2 \mathrm{d}\mathbf{x} 
	+ \sum_{k=1}^K \sum_{l>k}^K \beta_{k,l} \int_{\Gamma_{k,l}} h_{k,l}\left(u_k, u_l\right) \mathrm{d}\mathbf{x},
\end{align}
where $h_{k,l}\left(u_k, u_l\right)$ are coupling functions imposing a physics-conforming behavior of the combined solution along the interfaces $\Gamma_{k,l}$.
The form of the coupling functions is problem-dependent and must be chosen according to the physics governing the problem under investigation.
For example, assuming that the solution must be continuous along the interface $\Gamma_{k,l}$, an appropriate choice would be $h_{k,l}\left(u_k, u_l\right) = \left(u_k - u_l\right)^2$.

Implementation specifics are provided for the employed neural solvers in both test cases, in particular regarding \gls{ann} architecture (number of blocks as in figure \ref{fig:ritz_structure}, neurons per layer), learning rates, penalty terms, and training data inside the computational domain and on its boundary. We note that the reported values have been selected using a heuristic procedure, namely an initial grid search followed by manual fine-tuning, and that only the best-in-class configurations are presented. An extensive study with respect to optimizing \gls{ann} architecture and the hyperparameters, while definitely interesting and valuable, is out of scope for the present work. For the single-\gls{ann} neural solver, the architecture is always chosen such that the trainable parameters are slightly more than those of the multi-\gls{ann} neural solvers. With this choice we aim to underline the limited approximation capabilities of the single-\gls{ann} approach, even if the employed architecture is comparatively more expressive.
Finally, we note that the importance sampling scheme presented in section \ref{subsec:importance} is employed to discretize the energy functional in all cases, as the computational domains are constructed by means of \gls{cad} and a naive discretization of the energy functional leads to completely inaccurate results.

\subsection{Verification example: Dielectric cylinder in homogeneous electric field}
\label{subsec:cylinder}
\begin{figure}[t!]
	\begin{center}
		\begin{tikzpicture}
		\draw [very thick] (-5, -3) -- (-5,3);
		\draw [very thick] (5, -3) -- (5,3);
		\draw [dashed, very thick] (-5, -3) -- (5,-3);
		\draw [dashed, very thick] (-5, 3) -- (5,3);
		\node [anchor=east](n1) at (-5,0) {$\Gamma_\text{D}$};
		\node [anchor=west](n2) at (5,0) {$\Gamma_\text{D}$};
		\node [anchor=south](n3) at (0,3) {$\Gamma_\text{N}$};
		\node [anchor=north](n4) at (0,-3) {$\Gamma_\text{N}$};
		
		\draw[very thick, dotted, fill=gray, fill opacity=0.5] (0,0) circle (2);
		\node at (0,0)  {$\bullet$};
		\node at (0,1) {$\Omega_{\text{c}}$, $\epsilon_{\text{c}}$};
		\draw[thick, ->, >=stealth](0,0)--(-2^0.5,-2^0.5);
		\node [anchor=north west](n5) at (-0.5*2^0.5,-0.5*2^0.5){$r_0$};
		\node [anchor=north west](n6) at (-2.7,1){$\partial \Omega_\text{c}$};
		\node at (-3.5,2) {$\Omega_{\text{nc}}$, $\epsilon_{\text{nc}}$, $\mathbf{e}_\infty$};
		\draw [thick, ->, >=stealth] (6,-3)--(6,-2);
		\node [anchor=west] at (6,-2) {$y$};
		\draw [thick, ->, >=stealth] (6,-3)--(7,-3);
		\node [anchor=north] at (7,-3) {$x$};
		\draw(6,-3) circle (0.15 cm);
		\fill[black] (6,-3) circle (0.1cm);
		\end{tikzpicture}
		\caption{Two-dimensional cut of a dielectric cylinder suspended in a homogeneous electric field $\mathbf{e}_\infty$.}
		\label{fig:geom_cyl}
	\end{center}
\end{figure}
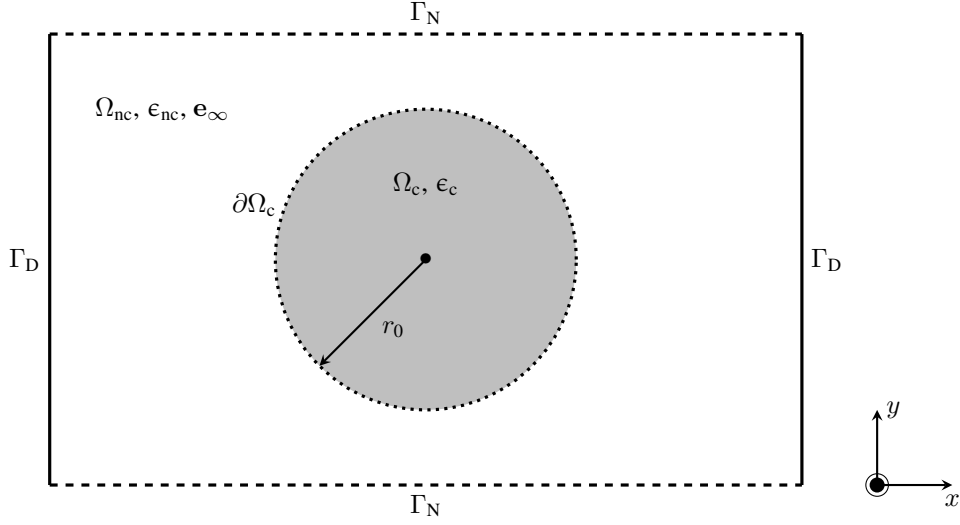

As a first test case, we consider the setting of an infinitely long dielectric cylinder suspended in a homogeneous electric field.
The geometry is shown in figure~\ref{fig:geom_cyl}, where the problem is reduced to a two-dimensional cut due to translational invariance along the $z$-axis. 
The computational domain is $\Omega = \left[-1,1\right] \times \left[-1,1\right]$ and the cylinder's domain is $\Omega_{\text{c}} = \left\{\mathbf{x} =\left(x,y\right) \:|\: \sqrt{x^2 + y^2} \leq r_0\right\}$, where $r_0 = 0.5$ is the radius of the cylinder.
The dielectric material of the cylinder has the relative scalar permittivity $\varepsilon_{\text{c}}=100$, while outside the cylinder the relative permittivity is $\varepsilon_\text{nc} = 1$. 
Hence, the computational domain consists of two subdomains with different material properties and the material interface coincides with the cylinder's boundary $\partial \Omega_\text{c}$.
The homogeneous electric field is given as $\mathbf{e}_\infty = \left(E_\infty, 0 \right)$, with $E_\infty=10$.
All sizes are expressed in SI units. 
We further assume Dirichlet \glspl{bc} on the left and right boundaries of $\Omega$, and Neumann \glspl{bc} on the top and bottom boundaries of $\Omega$.

The electric potential $u(\mathbf{x})$ in $\Omega$ can be computed by solving the \gls{bvp}
\begin{subequations}
	\label{eq:poisson_eq_gen}
	\begin{align}
		-\nabla \cdot \left(\varepsilon \left(\mathbf{x}\right)  \nabla u\left(\mathbf{x}\right)\right)&=0, && \mathbf{x}\in \Omega, \\
		u\left(\mathbf{x}\right)&=u^*\left(\mathbf{x}\right), &&\mathbf{x} \in \Gamma_{\text{D}}, \\
		\left(\nabla u\left(\mathbf{x}\right)\right)\cdot \mathbf{n}&=\left(\nabla u^*\left(\mathbf{x}\right)\right)\cdot \mathbf{n}, &&\mathbf{x}\in \Gamma_{\text{N}},
	\end{align}
\end{subequations}
where $\mathbf{n}$ denotes the outer normal unit vector, the permittivity $\varepsilon\left(\mathbf{x}\right)$ is given by
\begin{align}
	\varepsilon(\mathbf{x})=
	\begin{cases}
		\varepsilon_{\text{c}}, &\mathbf{x} \in \Omega_{\text{c}}, \\
		\varepsilon_{\text{nc}}, &\mathbf{x} \in \Omega \setminus \Omega_{\text{c}},
	\end{cases}
\end{align} 
and $u^*$, which is also the analytical solution to the problem, is given by
\begin{align}
	\label{eq:material_prob}
	u^{*}\left(\mathbf{x}\right)=-E_{\infty}x\:
	\begin{cases}
		1 - \frac{\varepsilon_{\text{c}} / \varepsilon_{\text{nc}} - 1}{\varepsilon_{\text{c}} / \varepsilon_{\text{nc}} + 1}\frac{r_0^2}{x^2 + y^2}, &\mathbf{x} \in \Omega \setminus \Omega_{\text{c}}, \\
		\frac{2}{\varepsilon_{\text{c}} / \varepsilon_{\text{nc}} + 1}, &\mathbf{x} \in \Omega_{\text{c}}.
	\end{cases}
\end{align}

\subsubsection{Energy functionals}
When no domain decomposition is employed, in which case a single \gls{ann} approximates the \gls{pde} solution over the whole domain, the energy functional to be minimized is
\begin{align}
	\label{eq:poisson_enfun}
	I(u) =&  \int_{\Omega_\text{c}} \frac{\varepsilon_{\text{c}}}{2} \left(\nabla u\right)^2 \mathrm{d}\mathbf{x} 
	+ \int_{\Omega \setminus \Omega_\text{c}} \frac{\varepsilon_{\text{nc}}}{2} \left(\nabla u\right)^2 \mathrm{d}\mathbf{x} 
	-  \int_{\Gamma_{\text{N}}} u \nabla u^* \cdot \mathbf{n} \, \mathrm{d}\mathbf{x} \nonumber \\ 
	&+ \beta_{\text{D}} \int_{\Gamma_{\text{D}}} \left(u - u^*\right)^{2}\mathrm{d}\mathbf{x}.
\end{align}
Using one \gls{ann} per subdomain in combination with the \gls{dg} formulation, the energy functional is modified to
\begin{align}
	\label{eq:poisson_enfun_DG}
	I\left(u\right) =&  \int_{\Omega_\text{c}} \frac{\varepsilon_{\text{c}}}{2} \left(\nabla u_{\text{c}}\right)^2 \mathrm{d}\mathbf{x} 
	+ \int_{\Omega \setminus \Omega_\text{c}} \frac{\varepsilon_{\text{nc}}}{2} \left(\nabla u_{\text{nc}}\right)^2 \mathrm{d}\mathbf{x} 
	-  \int_{\Gamma_{\text{N}}} u_{\text{nc}} \nabla u^* \cdot \mathbf{n} \, \mathrm{d}\mathbf{x} \nonumber \\ 
	&- \int_{\Gamma_{\text{D}}} \left\{\nabla u\right\}_{\Gamma}\left[u\right]_{\Gamma} \mathrm{d}\mathbf{x} 
	+ \frac{\beta_{\text{D}}}{2}\int_{\Gamma_{\text{D}}} \left[u-u^*\right]_{\Gamma}^2 \mathrm{d}\mathbf{x} \nonumber \\
	&- \int_{\partial \Omega_{\text{c}}} \left\{\nabla u\right\}_{\Gamma}\left[u\right]_{\Gamma}  \mathrm{d}\mathbf{x}
	+ \frac{\beta_{\text{i}}}{2}\int_{\partial \Omega_{\text{c}}}\left[u\right]_{\Gamma}^2 \mathrm{d}\mathbf{x},
\end{align}
where $u_{\text{c}}$, $u_{\text{nc}}$ denote the separate solutions in $\Omega_{\text{c}}$ and in $\Omega \setminus \Omega_{\text{c}}$, respectively.
Using one \gls{ann} per subdomain with coupling factors, we take advantage of the fact that the electric potential must remain continuous, thus, the corresponding energy functional is given by
\begin{align}
	\label{eq:poisson_enfun_physical}
	I(u) =&  \int_{\Omega_\text{c}} \frac{\varepsilon_{\text{c}}}{2} \left(\nabla u_\text{c}\right)^2 \mathrm{d}\mathbf{x} + \int_{\Omega \setminus \Omega_\text{c}} \frac{\varepsilon_{\text{nc}}}{2} \left( \nabla u_\text{nc}  \right)^2\mathrm{d}\mathbf{x} -  \int_{\Gamma_{\text{N}}} u_{\text{nc}} \left(\mathbf{x}\right) \nabla u^*\left(\mathbf{x}\right) \cdot \mathbf{n} \, \mathrm{d}\mathbf{x} \nonumber \\
	&+ \beta_{\text{D}} \int_{\Gamma_{\text{D}}} \left(u_\text{nc} - u^* \right)^2 \mathrm{d}\mathbf{x} 
	+ \beta_{\text{i}} \int_{\partial \Omega_{\text{c}}} \left(u_{\text{c}} - u_{\text{nc}} \right)^2 \mathrm{d}\mathbf{x}.
\end{align}

\subsubsection{Implementation specifics}
When no domain decomposition is employed, the single \gls{ann} consists of $6$ blocks (see figure~\ref{fig:ritz_structure}) with $10$ neurons per layer. 
For both domain decomposition-based neural solvers, the two \glspl{ann} consist of $4$ blocks with $10$ neurons per layer each.
For all approaches, we choose a learning rate of $\eta=1\cdot10^{-3}$ for $3\cdot10^4$ training epochs, followed by an additional $1\cdot10^4$ training epochs where the learning rate $\eta=1\cdot10^{-4}$ is used.
The penalty terms are $\beta_{\text{i}}=\beta_{\text{D}}=1\cdot 10^3$. 
The discretization of the energy functionals employs $4\cdot10^3$ sample points in $\Omega$, $1.2\cdot 10^3$ sample points on $\Gamma_{\text{D}}$ and $\Gamma_{\text{N}}$ and $6\cdot10^2$ sample points on $\partial \Omega_{\text{c}}$.


\subsubsection{Simulation results}
Instead of the electric potential, we focus on the static electric field $\mathbf{e} = -\nabla u$. 
Field continuity demands that, on the interface between the two materials, the tangential field component must remain continuous, i.e. $\mathbf{e}^{\text{t}}_{\text{c}} = \mathbf{e}^{\text{t}}_{\text{nc}}$, while the normal field component is discontinuous, i.e. $\mathbf{e}^{\text{n}}_{\text{c}} \neq \mathbf{e}^{\text{n}}_{\text{nc}}$.
In particular, it must hold that $\varepsilon_\text{c} \mathbf{e}^{\text{n}}_{\text{c}} = \varepsilon_\text{nc} \mathbf{e}^{\text{n}}_{\text{nc}}$, equivalently, $\varepsilon_\text{c} \nabla u_{\text{c}} \cdot \mathbf{n} = \varepsilon_\text{nc} \nabla u_{\text{nc}} \cdot \mathbf{n}$.

\begin{figure}[t!]
	\centering
	\includegraphics[width = \textwidth]{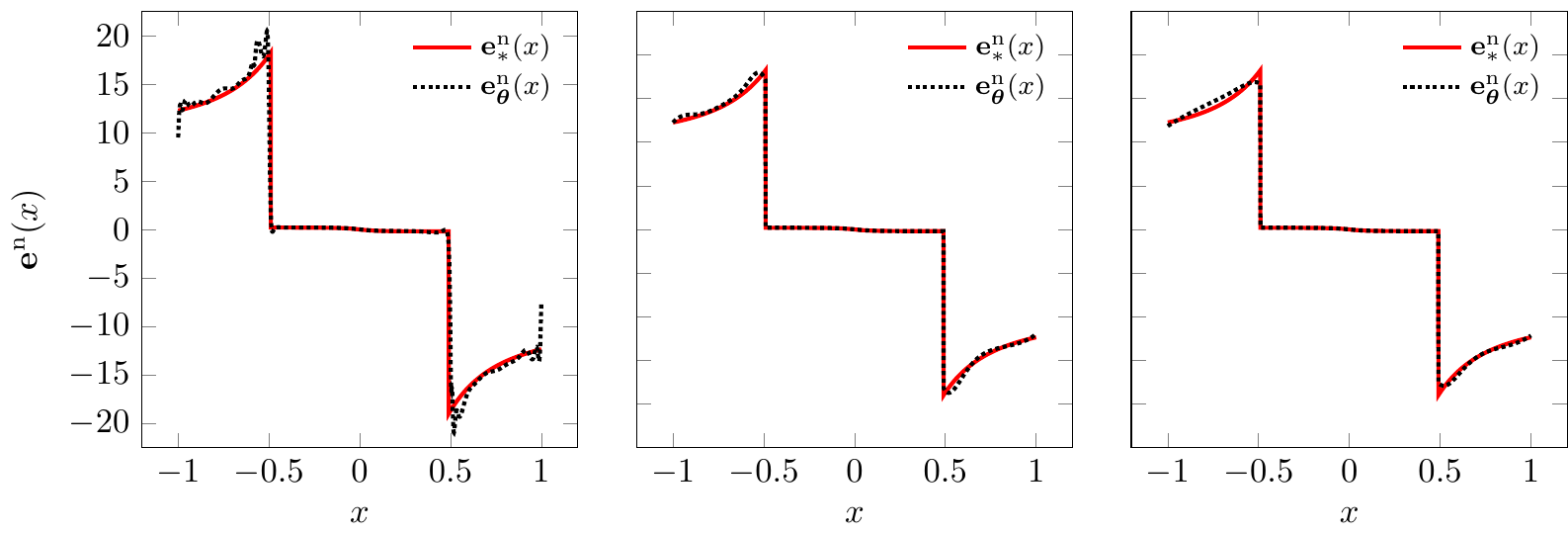}
	\caption{Normal component of the electric field along the line at $y=0.1$. Analytically computed field components are denoted with $*$, while ${\boldsymbol{\theta}}$ denotes field components computed with neural solvers. \textbf{Left:} Neural solver based on a single \gls{ann} without domain decomposition. \textbf{Middle:} Neural solver based on domain decomposition (two \glspl{ann}) and the \gls{dg} formulation. \textbf{Right:} Neural solver based on domain decomposition (two \glspl{ann}) and coupling factors.}
	\label{fig:normal_components}
\end{figure}

Comparisons between \gls{ann}-based normal field components $\mathbf{e}^{\text{n}}_{\boldsymbol{\theta}}$ and the analytical one $\mathbf{e}^{\text{n}}_{*}$ for each neural solver are presented in figure \ref{fig:normal_components}. 
For a better illustration of the results, the normal electric field component is evaluated along the line $y=0.1$ that cuts the domain $\Omega$ horizontally.
As can be observed, a single-\gls{ann} cannot produce a physics-conforming solution and results in Gibbs' overshooting phenomena that occur exactly on the interface between the two subdomains, i.e. exactly where the material discontinuity exists.
The single-\gls{ann} seems to have severe approximation difficulties close to the domain boundaries at $x=-1$ and $x=1$ as well.
On the contrary, both domain decomposition-based neural solvers produce physics-conforming solutions that accurately capture the behavior of the electric field's normal component in view of the material discontinuity, as well as close to the domain boundaries.
Small approximation difficulties in between the domain boundaries and the material interface can be observed for both methods, however, with evident differences.  Given that both neural solvers employ \glspl{ann} with the exact same architecture and hyperparameter values, this result can only be attributed to the different loss functions and their impact upon the training process.

\subsection{Engineering application: Electric machine simulation}
\label{subsec:pmsm}
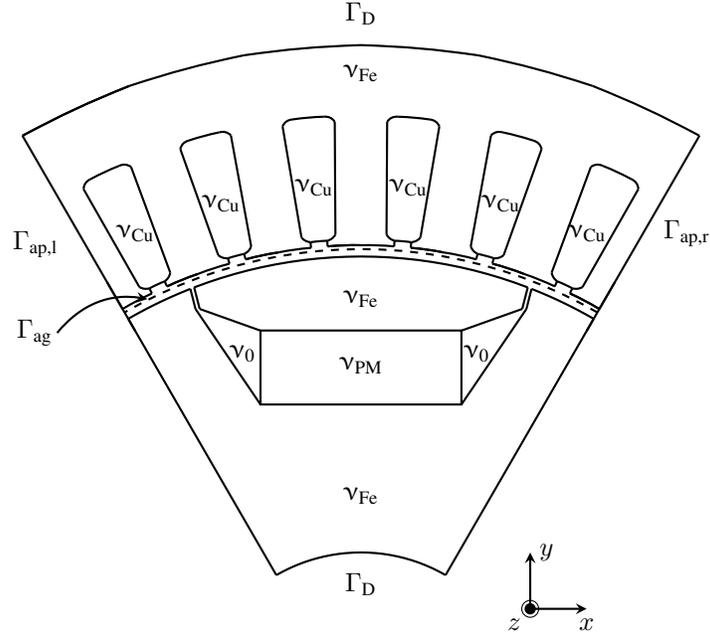
\begin{figure}[!t]
\centering
\begin{tikzpicture}[scale=0.75]

\draw [black,thick,domain=120-8.5:120] plot ({3*2.75*cos(\x)}, {3*2.75*sin(\x)});
\draw [black,thick,domain=60:60+8.5] plot ({3*2.75*cos(\x)}, {3*2.75*sin(\x)});
\draw [black,thick,domain=90-21:90+21] plot ({3*2.75*cos(\x)}, {3*2.75*sin(\x)});
\draw [black,thick,dashed, domain=60:120] plot ({3*2.79375*cos(\x)}, {3*2.79375*sin(\x)});
     
\draw [black,thick, domain=60:120] plot ({3*cos(\x)}, {3*sin(\x)});
\draw[thick] ({3*cos(60)},{3*sin(60)}) -- ({3*2.75*cos(60)},{3*2.75*sin(60)}); 
\draw[thick] ({3*cos(120)},{3*sin(120)}) -- ({3*2.75*cos(120)},{3*2.75*sin(120)}); 

\draw[thick] ({3*1.9668*cos(107.57},{3*1.9668*sin(107.57)}) rectangle ({3*2.3875*cos(75.6)},{3*2.3875*sin(75.6)});

\draw[thick] ({3*1.9668*cos(107.57)},{3*1.9668*sin(107.57)}) -- ({4*1.9668*cos(120-8.5)-0.05},{4*1.9668*sin(120-8.5)});
\draw[thick] ({4*1.9668*cos(120-8.5)-0.05},{4*1.9668*sin(120-8.5)}) -- ({3*2.75*cos(120-8.5)}, {3*2.75*sin(120-8.5)});
\draw[thick] ({4*1.9668*cos(120-8.5)+0.025},{4*1.9668*sin(120-8.5)}) -- ({3*2.75*cos(90+21)}, {3*2.75*sin(90+21)}); 
\draw[thick] ({3*1.9668*cos(107.57},{3*1.9668*sin(107.57)+1.31}) -- ({4*1.9668*cos(120-8.5)+0.025},{4*1.9668*sin(120-8.5)});

\draw[thick] ({-3*1.9668*cos(107.57)},{3*1.9668*sin(107.57)}) -- ({-4*1.9668*cos(120-8.5)+0.05},{4*1.9668*sin(120-8.5)});
\draw[thick] ({-4*1.9668*cos(120-8.5)+0.05},{4*1.9668*sin(120-8.5)}) -- ({-3*2.75*cos(120-8.5)}, {3*2.75*sin(120-8.5)});
\draw[thick] ({-4*1.9668*cos(120-8.5)-0.025},{4*1.9668*sin(120-8.5)}) -- ({-3*2.75*cos(90+21)}, {3*2.75*sin(90+21)}); 
\draw[thick] ({-3*1.9668*cos(107.57},{3*1.9668*sin(107.57)+1.31}) -- ({-4*1.9668*cos(120-8.5)-0.025},{4*1.9668*sin(120-8.5)});

\draw ({3*2.3875*sin(75.6)*cos(90)}, {3*1.9668*sin(107.57)+0.7}) node [anchor=center]{$\nu_{\text{PM}}$};
\draw (0, 4) node [anchor=center]{$\nu_{\text{Fe}}$};
\draw (0, 7.5) node [anchor=center]{$\nu_{\text{Fe}}$};
\draw ({3*2.3875*sin(75.6)*cos(90)-2.1}, {3*1.9668*sin(107.57)+0.9}) node [anchor=center]{$\nu_{0}$};
\draw ({3*2.3875*sin(75.6)*cos(90)+2.05}, {3*1.9668*sin(107.57)+0.9}) node [anchor=center]{$\nu_{0}$};

\draw [thick, ->, >=stealth] (3,2)--(3,3);
\draw(3,2) circle (0.15 cm);
\fill[black] (3,2) circle (0.1cm);
\draw (3,2) node [anchor=north east]{$z$};
\node [anchor=west] at (3,3) {$y$};
\draw [thick, ->, >=stealth] (3,2)--(4,2);
\node [anchor=north] at (4,2) {$x$};
	
\tikzmath{\ri=8.45; \ro=12;\del=0.1;}
\draw[thick]({\ri*cos(120)}, {\ri*sin(120)}) to ({\ro*cos(120)}, {\ro*sin(120)});
\draw[thick]({\ri*cos(60)}, {\ri*sin(60)}) to ({\ro*cos(60)}, {\ro*sin(60)});

\draw[thick]({\ro*cos(120)}, {\ro*sin(120)}) to[bend left=3] ({\ro*cos(110)}, {\ro*sin(110)});

\foreach \ai in {0,10,20,30,40,50}{
 	
 	\draw[thick]({\ro*cos(120-\ai)}, {\ro*sin(120-\ai)}) to[bend left=3] ({\ro*cos(110-\ai)}, {\ro*sin(110-\ai)});
	
	\draw[thick]({\ri*cos(120-\ai)}, {\ri*sin(120-\ai)}) to[bend left=3] ({\ri*cos(120-4-\ai)}, {\ri*sin(120-4-\ai)});

	\draw[thick]({\ri*cos(120-6-\ai)}, {\ri*sin(120-6-\ai)}) to[bend left=3] ({\ri*cos(120-10-\ai)}, {\ri*sin(120-10-\ai)});
		
	\draw[thick]({\ri*cos(120-6-\ai)}, {\ri*sin(120-6-\ai)}) to[bend left=3] ({\ri*cos(120-10-\ai)}, {\ri*sin(120-10-\ai)});

	\draw[thick]({(\ri+\del)*cos(120-4-\ai)}, {(\ri+\del)*sin(120-4-\ai)}) to[bend left=3] ({(\ri+\del)*cos(120-6-\ai)}, {(\ri+\del)*sin(120-6-\ai)});

	
	\draw[thick]({(\ri+23*\del)*cos(120-3.5-\ai)}, {(\ri+23*\del)*sin(120-3.5-\ai)}) to[bend left=3] ({(\ri+23*\del)*cos(120-6.5-\ai)}, {(\ri+23*\del)*sin(120-6.5-\ai)});


	\draw[thick]({(\ri+\del)*cos(120-4-\ai)}, {(\ri+\del)*sin(120-4-\ai)}) to[bend left=20] ({(\ri+2*\del)*cos(120-3-\ai)}, {(\ri+2*\del)*sin(120-3-\ai)});

	\draw[thick]({(\ri+2*\del)*cos(120-3-\ai)}, {(\ri+2*\del)*sin(120-3-\ai)}) to ({(\ri+22*\del)*cos(120-2.5-\ai)}, {(\ri+22*\del)*sin(120-2.5-\ai)});

	\draw[thick]({(\ri+22*\del)*cos(120-2.5-\ai)}, {(\ri+22*\del)*sin(120-2.5-\ai)}) to[bend left=30] ({(\ri+23*\del)*cos(120-3.5-\ai)}, {(\ri+23*\del)*sin(120-3.5-\ai)});

	\draw[thick]({(\ri+\del)*cos(120-6-\ai)}, {(\ri+\del)*sin(120-6-\ai)}) to[bend right=20] ({(\ri+2*\del)*cos(120-7-\ai)}, {(\ri+2*\del)*sin(120-7-\ai)});

	\draw[thick]({(\ri+2*\del)*cos(120-7-\ai)}, {(\ri+2*\del)*sin(120-7-\ai)}) to ({(\ri+22*\del)*cos(120-7.5-\ai)}, {(\ri+22*\del)*sin(120-7.5-\ai)});

	\draw[thick]({(\ri+22*\del)*cos(120-7.5-\ai)}, {(\ri+22*\del)*sin(120-7.5-\ai)}) to[bend right=30] ({(\ri+23*\del)*cos(120-6.5-\ai)}, {(\ri+23*\del)*sin(120-6.5-\ai)});

	\draw[thick]({\ri*cos(120-4-\ai)}, {\ri*sin(120-4-\ai)}) to[bend left=3] ({(\ri+\del)*cos(120-4-\ai)}, {(\ri+\del)*sin(120-4-\ai)});

	\draw[thick]({\ri*cos(120-6-\ai)}, {\ri*sin(120-6-\ai)}) to[bend left=3] ({(\ri+\del)*cos(120-6-\ai)}, {(\ri+\del)*sin(120-6-\ai)});
};

\draw (0, 11.5) node [anchor=center]{$\nu_{\text{Fe}}$};

\draw ({9*cos(115)-0.2}, {9*sin(115)+0.5}) node [anchor=center]{$\nu_{\text{Cu}}$};
\draw ({9*cos(105)-0.15}, {9*sin(105)+0.5}) node [anchor=center]{$\nu_{\text{Cu}}$};
\draw ({9*cos(95)-0.05}, {9*sin(95)+0.5}) node [anchor=center]{$\nu_{\text{Cu}}$};
\draw ({9*cos(85)+0.05}, {9*sin(85)+0.5}) node [anchor=center]{$\nu_{\text{Cu}}$};
\draw ({9*cos(75)+0.15}, {9*sin(75)+0.5}) node [anchor=center]{$\nu_{\text{Cu}}$};	
\draw ({9*cos(65)+0.2}, {9*sin(65)+0.5}) node [anchor=center]{$\nu_{\text{Cu}}$};	

\draw ({10*cos(120)-0.2},{10*sin(120)-0.2})node [anchor=east]{$\Gamma_\text{ap,l}$};
\draw ({10*cos(60)+0.2},{10*sin(60)})node [anchor=west]{$\Gamma_\text{ap,r}$};

\draw (0,12+0.2) node [anchor=south]{$\Gamma_\text{D}$};
\draw (0,3-0.2) node [anchor=north]{$\Gamma_\text{D}$};

\draw[thick] ({8*cos(120)},{8*sin(120)})--({9*cos(120)},{9*sin(120)});
\draw[thick] ({8*cos(60)},{8*sin(60)})--({9*cos(60)},{9*sin(60)});

\draw ({9*cos(130)}, {9*sin(130)}) node [anchor=center]{$\Gamma_{\text{ag}}$};
\draw[thick, ->, >=stealth]({9*cos(130)+0.4}, {9*sin(130)}) to[bend left=30] ({9*cos(130)+2}, {9*sin(130)+0.6});



\end{tikzpicture}
\caption{One-sixth of the \gls{pmsm}'s geometry in two dimensions.}
\label{fig:motor_seg}
\end{figure}

We now consider a real-world engineering application, namely the simulation of a six-pole \gls{pmsm} \citep{bontinck2018isogeometric, bontinck2018robust}.
An illustration is provided in figure~\ref{fig:motor_seg}, where only one-sixth of a two-dimensional cross-section of the \gls{pmsm} is considered, due to rotational symmetry and translational invariance along the $z$-axis.
The \gls{pmsm} consists of stator and a rotor, which are separated by an air gap. 
The rotor consists of an iron part, a permanent magnet, and two air slots connected to the air gap.
The stator consists of an iron part and six slots where the copper windings carrying the source current are placed. 
The air gap is separated into a rotor part and a stator part by the interface $\Gamma_\text{ag}$ shown in figure~\ref{fig:motor_seg}.
The different material domains are distinguished by the respective magnetic reluctivities, denoted with $\nu_{\text{Fe}}$ for iron, $\nu_{\text{Cu}}$ for copper, $\nu_{\text{PM}}$ for the permanent magnet, and $\nu_0$ for air.
The geometry of the \gls{pmsm} is originally given in $90$ non-overlapping patches, 12 for the rotor and 78 for the stator.  
All patches are described by means of \gls{nurbs} \citep{bhat2018modelling, merkel2021shape}.
A detailed presentation of the geometrical and material parameters of the \gls{pmsm} is available in \ref{sec:appendix}.

The magnetic field distribution on the \gls{pmsm} is obtained by the magnetostatic formulation 
\begin{equation}
	\label{eq:pmsm_mag_model}
	\nabla \times \left(\nu \nabla \times \mathbf{a}\right)=\mathbf{j}_{\text{src}}+\nabla \times \mathbf{m},
\end{equation}
where $\nu = \nu(\mathbf{x}) = \nu(x,y)$ is a piece-wise constant magnetic reluctivity, $\mathbf{a}$ the magnetic vector potential, $\mathbf{j}_{\text{src}}$ the source current density, and $\mathbf{m}=\left(m_x,m_y\right)$ the magnetization vector of the permanent magnet.
We assume that the machine operates in generator mode under no load condition, such that $\mathbf{j}_{\text{src}}=0$. 
We additionally assume a constant magnetization in the $y$-direction, given as $\mathbf{m} = \left(0, \nu_0 B_{\text{r}}\right)$, such that the remanent magnetic flux density of the permanent magnet is $\mathbf{b}_{\text{r,PM}} = \left(0, B_{\text{r}}\right)$.
Under these assumptions, the solution to \eqref{eq:pmsm_mag_model} has only a $z$-component, i.e. $\mathbf{a} = \left(0, 0, u\left(\mathbf{x}\right)\right) = \left(0, 0, u\left(x,y\right)\right)$.
We choose anti-periodic \glspl{bc} on the left and right boundaries of the computational domain and homogeneous Dirichlet \glspl{bc} on the upper and lower boundaries. We denote the anti-periodic boundary on the left and right hand side of the machine with $\Gamma_{\text{ap, l}}$ and $\Gamma_{\text{ap, r}}$ and the Dirichlet boundary with $\Gamma_{\text{D}}$.

\subsubsection{Energy functionals}
The assumptions of the \gls{pmsm} model reduce the \gls{pde} \eqref{eq:pmsm_mag_model} to a scalar Poisson equation, the energy functional of which is given as
\begin{align}
	\label{eq:en_func_pmsm}
	I(u)=\frac{1}{2} \int_{\Omega}\nu \, |\nabla u|^2 \, \mathrm{d}\mathbf{x}-\int_{D_{\text{PM}}}\begin{pmatrix} -m_y \\ m_x\end{pmatrix}\cdot\nabla u \, \mathrm{d}\mathbf{x},
\end{align}
where $D_{\text{PM}}$ denotes the domain of the permanent magnet placed in the rotor.
Not considering any domain decomposition, the energy functional on the \gls{pmsm} is given by 
\begin{align}
	\label{eq:en_func_sANN}
	I(u)=&\frac{1}{2} \int_{\Omega} \nu |\nabla u|^2 \mathrm{d}\mathbf{x} +  \nu_0 B_\text{r}  \int_{D_{\text{PM}}}\frac{\partial u}{\partial x} \mathrm{d}\mathbf{x}  \nonumber  \\ 
	& + \beta_{\text{D}} \int_{\Gamma_{\text{D}}} u^2  \mathrm{d}\mathbf{x} + \beta_{\text{ap,l}}\int_{\Gamma_{\text{ap, l}}} \left(u(x,y) +u(-x,y) \right)^2 \mathrm{d}\mathbf{x}.
\end{align}
Note that by sampling the anti-periodic boundaries symmetrically, it is sufficient to impose the anti-periodic \glspl{bc} by integrating along either $\Gamma_{\text{ap, l}}$ or $\Gamma_{\text{ap, r}}$, where the former is chosen in \eqref{eq:en_func_sANN}.
Turning to the decomposed domain, the \gls{dg} formulation results in the energy functional
\begin{align}
	\label{eq:en_func_dg}
	I(u)=&\frac{1}{2} \sum_{k=1}^{K} \int_{D_k} \nu_k |\nabla u_k|^2  \, \mathrm{d}\mathbf{x} + \nu_0 B_\text{r} \int_{D_{\text{PM}}}\frac{\partial u}{\partial x}  \, \mathrm{d}\mathbf{x} \nonumber \\  &
	- \sum_{k=1}^{K} \sum_{l>k}^{K} \int_{\Gamma_{k,l}} \left\{\nabla u_k \right\}_{\Gamma_{k,l}}  \left[u_k\right]_{\Gamma_{k,l}} \mathrm{d}\mathbf{x} 
	+ \sum_{k=1}^{K} \sum_{l>k}^{K} \frac{\beta_{k,l}}{2} \int_{\Gamma_{k,l}} \left[u\right]_{\Gamma_{k,l}}^2 \mathrm{d}\mathbf{x} \nonumber \\  &
	- \int_{\Gamma_{\text{D}}} \left\{\nabla u \right\}_{\Gamma_{\text{D}}} \left[u\right]_{\Gamma_{\text{D}}} \mathrm{d}\mathbf{x} 
	+ \frac{\beta_{\text{D}}}{2} \int_{\Gamma_{\text{D}}} \left[u \right]_{\Gamma_{\text{D}}}^2 \mathrm{d}\mathbf{x}\\ &
	- \int_{\Gamma_{\text{ap, l}}} \left\{\nabla u \right\}_{\Gamma_{\text{ap, l}}}  \left[u\right]_{\Gamma_{\text{ap, l}}} \mathrm{d}\mathbf{x}\nonumber
	+ \frac{\beta_{\text{ap,l}}}{2} \int_{\Gamma_{\text{ap, l}}} \left[u\right]_{\Gamma_{\text{ap, l}}}^2 \mathrm{d}\mathbf{x},
\end{align}
where the average and jump operators on the left anti-periodic boundary $\Gamma_{\text{ap, l}}$ are defined as $\left\{\nabla u \right\}_{\Gamma_{\text{ap, l}}} =\frac{1}{2} \left(\nabla u (x,y)+ \nabla u(-x,y) \right)$ and $ \left[u\right]_{\Gamma_{\text{ap, l}}}:=\left( u(x,y)+u(-x,y) \right) \cdot \mathbf{n}_{\text{ap,l}}$, respectively.
When coupling factors are employed, the corresponding energy functional reads
\begin{align}
	\label{eq:en_func_pi}
	I(u)=&\frac{1}{2} \sum_{k=1}^{K} \int_{D_k} \nu_k |\nabla u_k|^2  \, \mathrm{d}\mathbf{x} + \nu_0 B_\text{r}  \int_{D_{\text{PM}}}\frac{\partial u}{\partial x}  \, \mathrm{d}\mathbf{x}  \nonumber \\  
	& + \sum_{k=1}^{K} \sum_{l>k}^{K} \beta_{k,l} \int_{\Gamma_{k,l}} h_{k,l}\left(u_k, u_l\right) \mathrm{d}\mathbf{x} +\beta_{\text{D}} \int_{\Gamma_{\text{D}}} u\left(\mathbf{x}\right)^2  \mathrm{d}\mathbf{x}   \nonumber \\ 
	& + \beta_{\text{ap,l}}\int_{\Gamma_{\text{ap, l}}} \left(u(x,y) +u(-x,y) \right)^2 \mathrm{d}\mathbf{x},
\end{align}
where $h_{k,l}\left(u_k, u_l \right) = \left(u_k - u_l \right)^2$, taking advantage of the continuity of the magnetic vector potential.

\subsubsection{Implementation specifics}
\begin{figure}[t!]
	\centering
	\includegraphics[width=1\textwidth, height=0.7\textwidth]{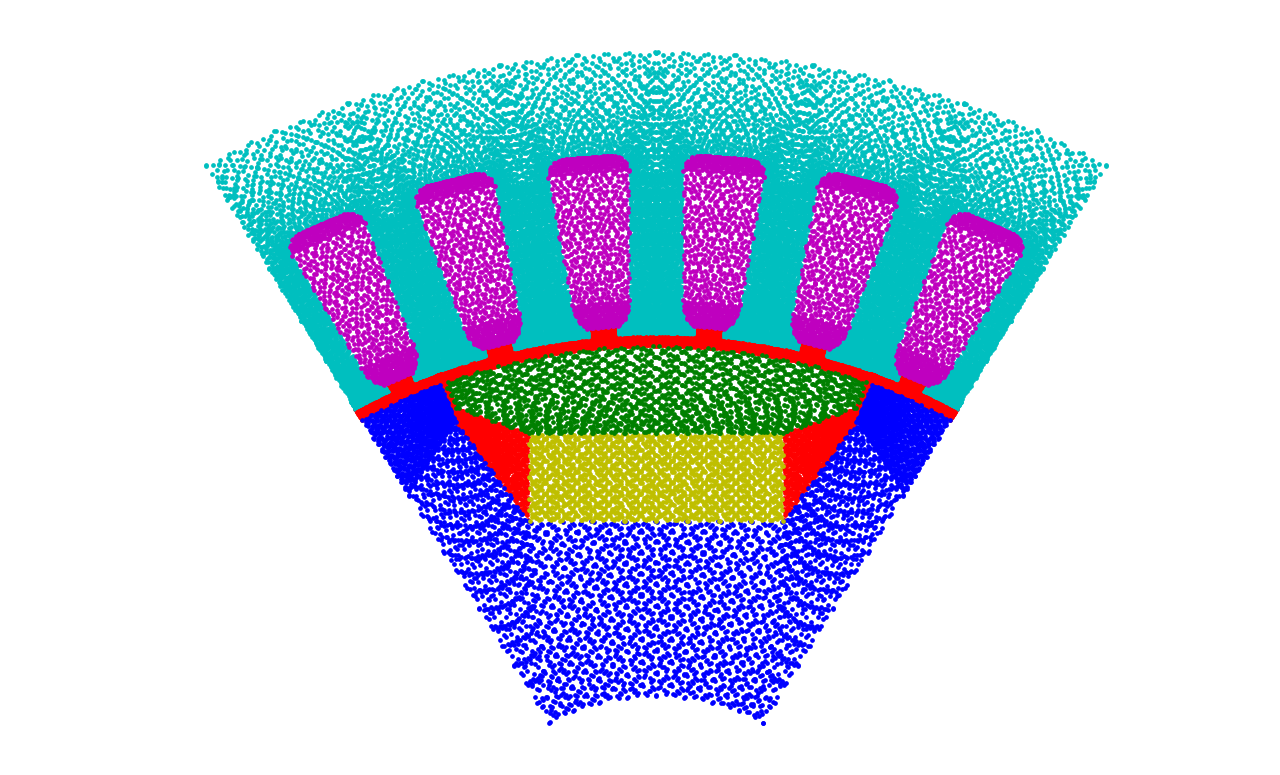}
	\caption{Domain partitioning and \gls{qmc} sampling on the geometry of the \gls{pmsm}.}
	\label{fig:MC_samples_mach}
\end{figure}

When no domain decomposition is considered, the single \gls{ann} consists of $30$ blocks with $24$ neurons per hidden layer, thus resulting in $36025$ trainable parameters.
In the case of the domain decomposition-based neural solvers, instead of using the original patches that form the geometry of the \gls{pmsm}, we construct subdomains by combining patches with the same material property and allocate a single \gls{ann} for each.
This yields a material-based domain decomposition which is illustrated in figure \ref{fig:MC_samples_mach} and further explained in table~\ref{tab:col_cod}.
Table~\ref{tab:col_cod} also provides details regarding the architecture of the \gls{ann} assigned to each subdomain, as well as the number of trainable parameters $|\boldsymbol{\theta}|$ and the size of the training data set $N_\text{sd}$ per subdomain. 
The total number of trainable parameters for the \gls{dg}-based neural solver is $33354$.
The training data set consists of $63 \cdot 10^3$ sampling points, the distribution of which on the domain of the \gls{pmsm} is shown in figure \ref{fig:MC_samples_mach}. 
For the implementation of the boundary and interface conditions, we use $2800$ sampling points on the anti-periodic boundaries, $1100$ sampling points on the Dirichlet boundaries, and $13500$ sampling points on the subdomain interfaces.
The \glspl{ann} are trained with the learning rate $\eta=1\cdot 10^{-3}$ for $5000$ epochs.
Last, the penalty factors $\beta_{\text{D}}$, $\beta_{k,l}$, and $\beta_{\text{ap, l}}$ in \eqref{eq:en_func_dg} are equal to either $\beta_{\text{rt}} = 1 \cdot 10^{5}$ or $\beta_{\text{st}} = 2 \cdot 10^{4}$, depending on whether the original patch belongs to the rotor or the stator, respectively.

\begin{table}[b!]
	\centering
	\begin{tabular}{c c c c c c c}
		\hline\hline
		&Subdomain & Color in Fig.~\ref{fig:MC_samples_mach} & Blocks & Neurons per Layer & $|\boldsymbol{\theta}|$ & $N_\text{sd} / 10^3$ \\ [0.5ex] 
		\hline
		\\[0.25ex]
		&Outer rotor yoke & Green & $8$ & 15 & $3872$ & $2$  \\[0.25ex]
		&Inner rotor yoke & Blue & $8$ & 15 & $3872$  & $6$ \\
		&Air Slits/Air Gap & Red & $15$ & 15 &  $7246$   & $26$ \\
		&Permanent Magnet & Yellow & $8$ & 15 & $3872$  & $2$ \\
		&Stator Yoke & Cyan & $15$ & 15 &  $7246$ & $21$ \\[0.25ex]
		&Stator Windings & Magenta & $15$ & 15 & $7246$  & $6$\\
		\\
		\hline
		\hline
	\end{tabular}
	\caption{Domain partitioning and \gls{ann} architecture for the domain decomposition-based neural solvers.}
	\label{tab:col_cod}
\end{table}  

\subsubsection{Simulation results} 

%

\begin{figure}[t!]
	\centering
	\begin{subfigure}[b]{.49\textwidth}
		\centering
		\includegraphics[width=1\textwidth, height=0.85\textwidth]{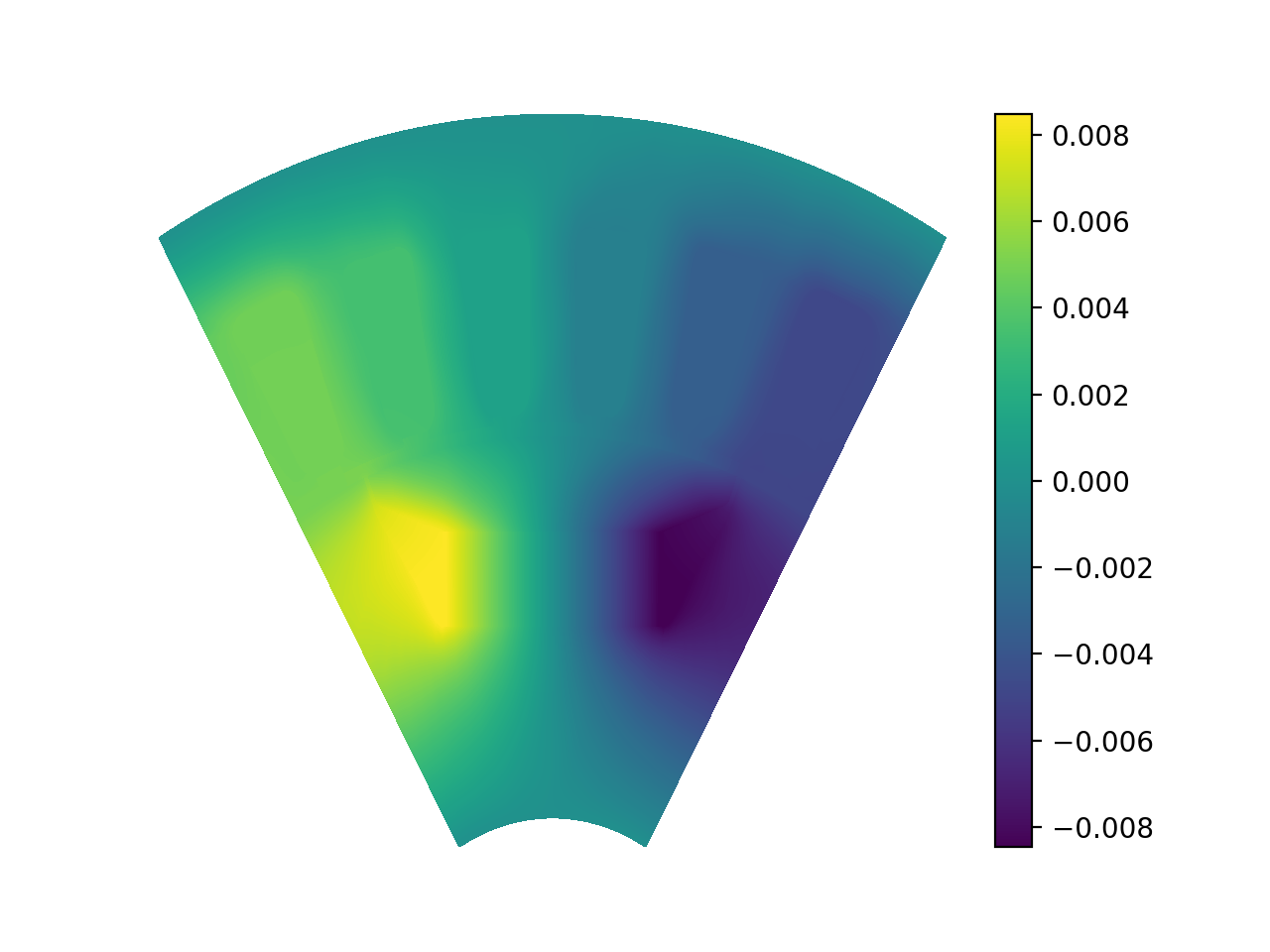}
		\caption{\gls{dg}-based neural solver.}
	\end{subfigure}%
	\begin{subfigure}[b]{.49\textwidth}
		\centering
		\includegraphics[width=1\textwidth, height=0.85\textwidth]{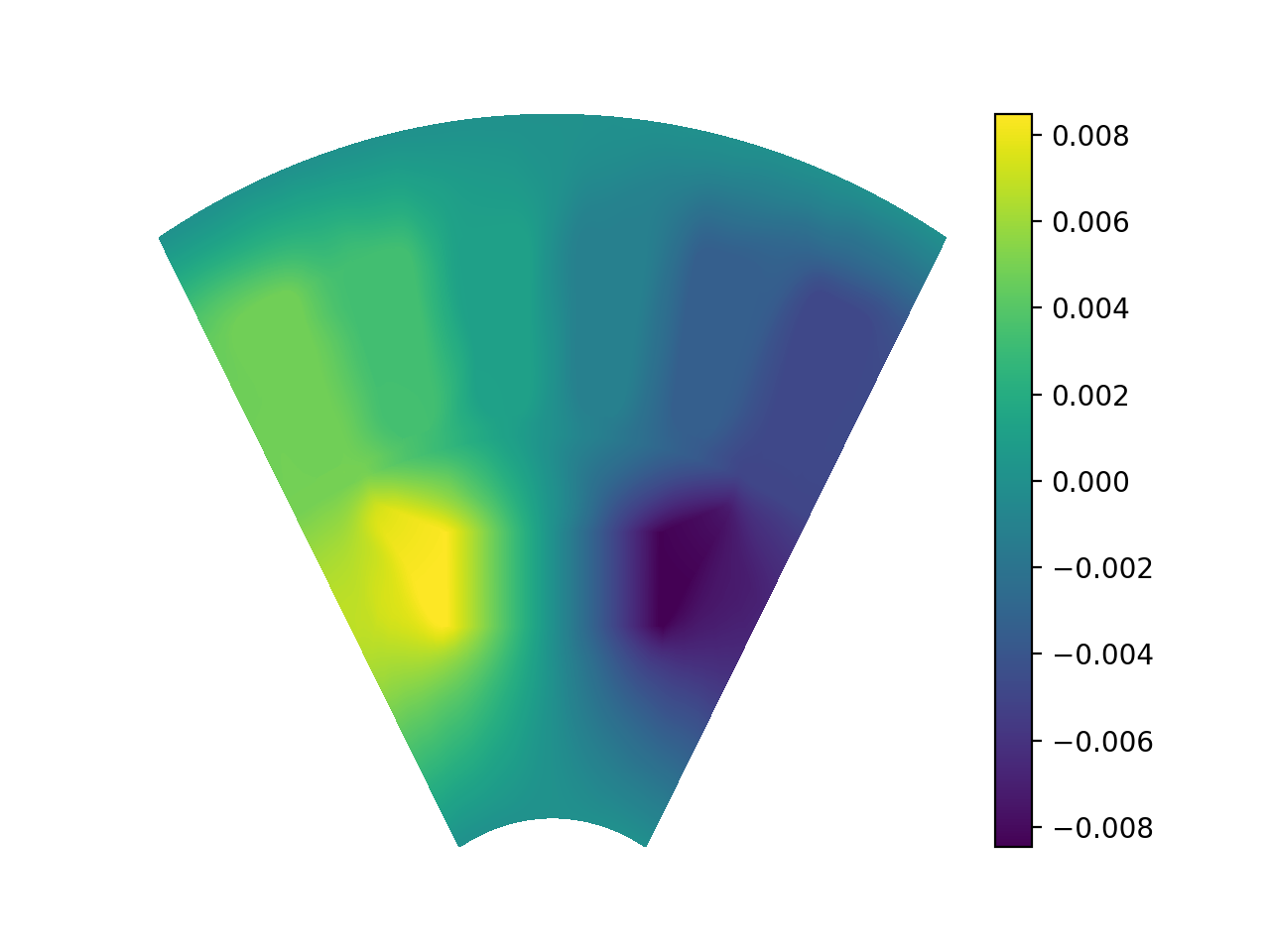}
		\caption{Coupling factors-based neural solver.}
	\end{subfigure}%
	\\[0.5em]
	\begin{subfigure}[b]{.49\textwidth}
		\centering
		\includegraphics[width=1\textwidth, height=0.85\textwidth]{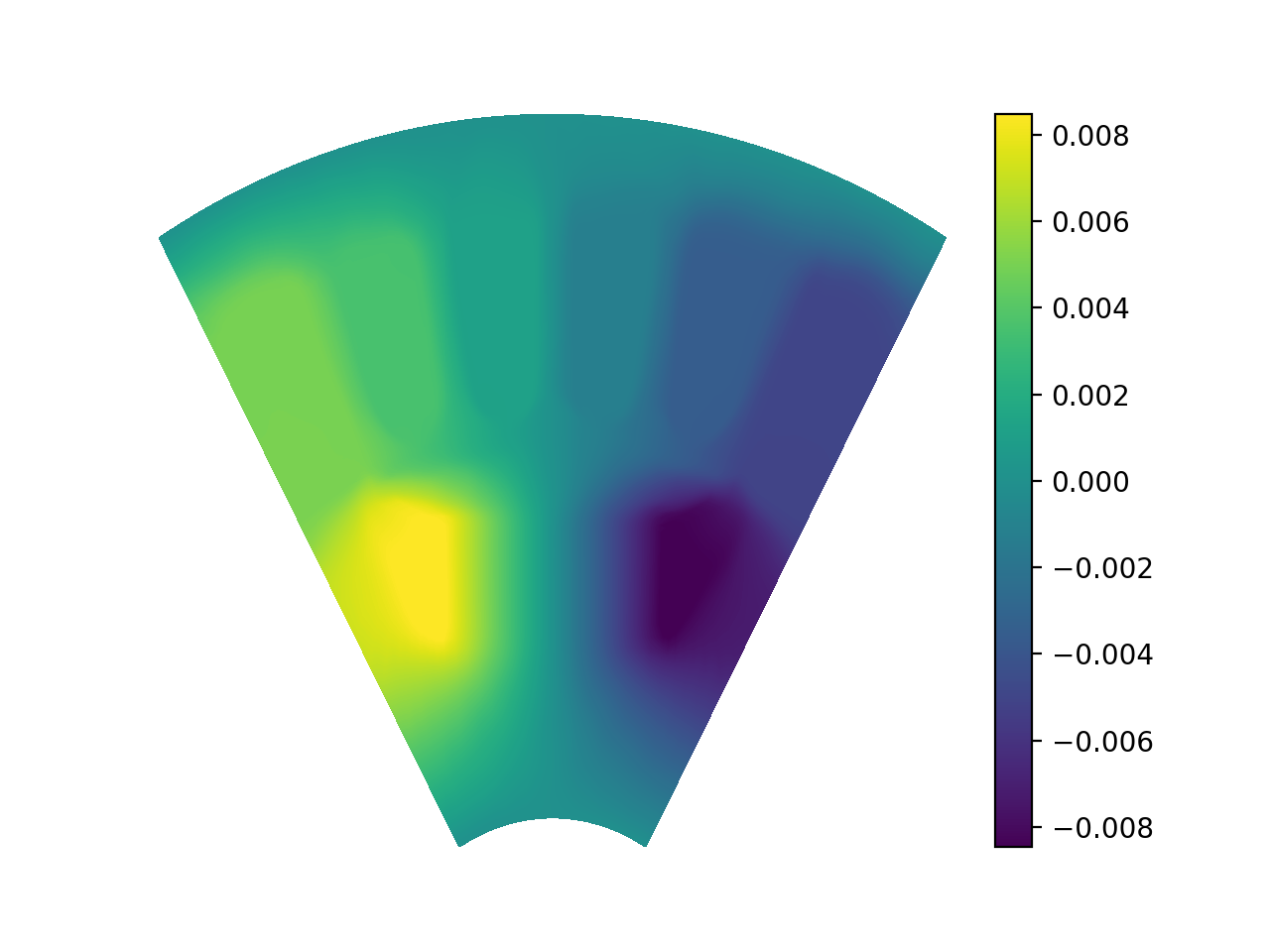}
		\caption{Single \gls{ann}-based neural solver.}
	\end{subfigure}%
	\begin{subfigure}[b]{.49\textwidth}
		\centering
		\includegraphics[width=1\textwidth, height=0.85\textwidth]{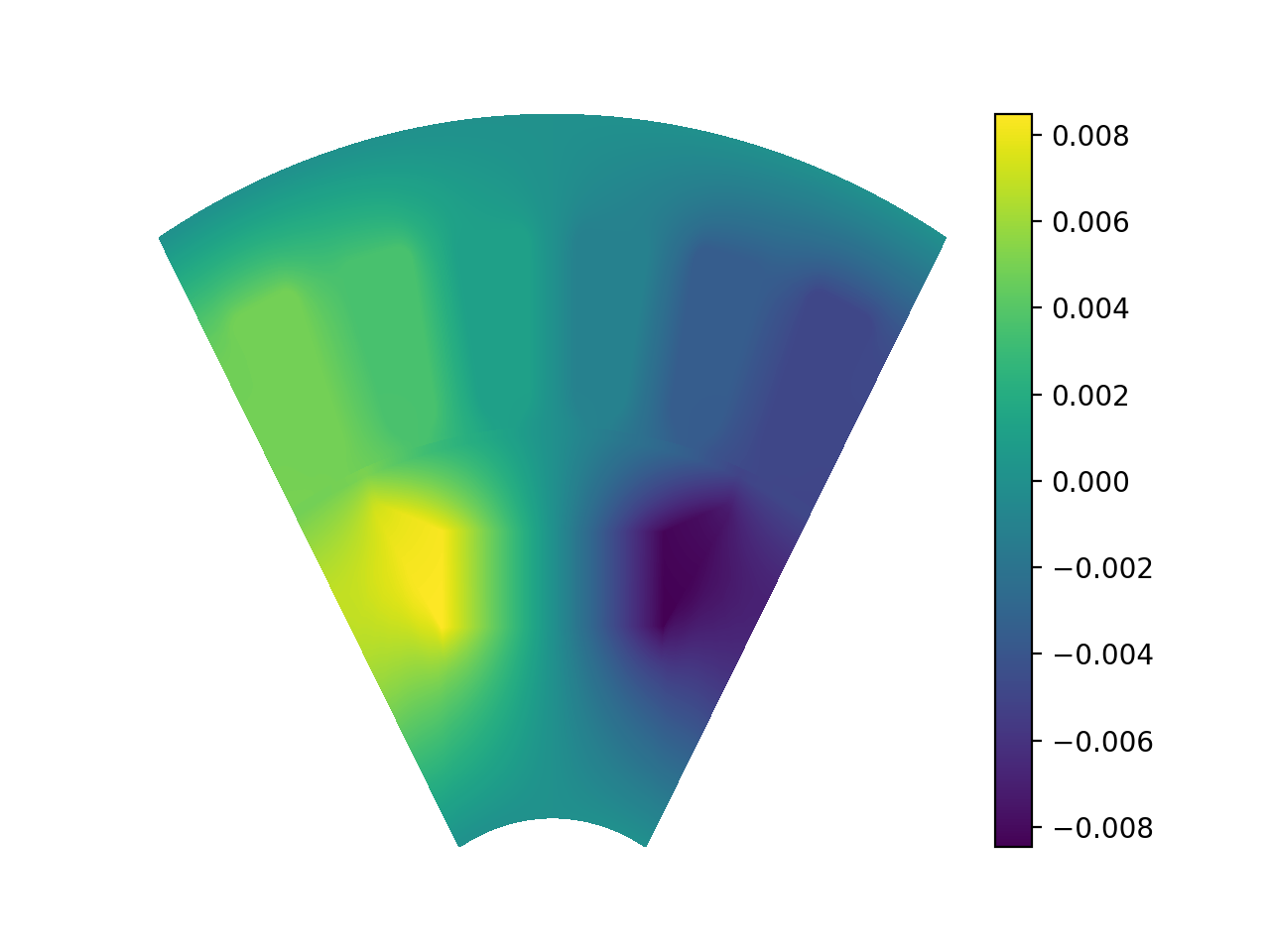}
		\caption{\gls{iga}-\gls{fem} solver.}
	\end{subfigure}%
	\caption{Solutions of the magnetic vector potential $\mathbf{a}_z$ $\left[\si{V  s m^{-1}}\right]$ using the reference \gls{iga}-\gls{fem} solver, the neural solver based on a single \gls{ann} without domain decomposition, the neural solver utilizing domain decomposition based on the \gls{dg} formulation, and the neural solver utilizing domain decomposition based on coupling factors.}
	\label{fig:DRM_sols}
\end{figure}

Figure~\ref{fig:DRM_sols} shows the magnetic vector potential solutions provided by the three employed neural solvers at the end of the training process, as well as a reference solution obtained with an \gls{iga}-\gls{fem} solver \citep{bontinck2018isogeometric}.
It is visually evident that the solution provided by a single \gls{ann} differs significantly from the reference, particularly in the regions on the left and right of the permanent magnet (bright yellow and deep blue regions). 
On the contrary, the solutions provided by both domain decomposition-based neural solvers, i.e. either based on the \gls{dg} formulation or on coupling factors, are visually almost identical to the reference.

\begin{figure}[t!]
	\centering
	\begin{subfigure}[b]{.49\textwidth}
		\centering
		\includegraphics[width=1\textwidth, height=0.85\textwidth]{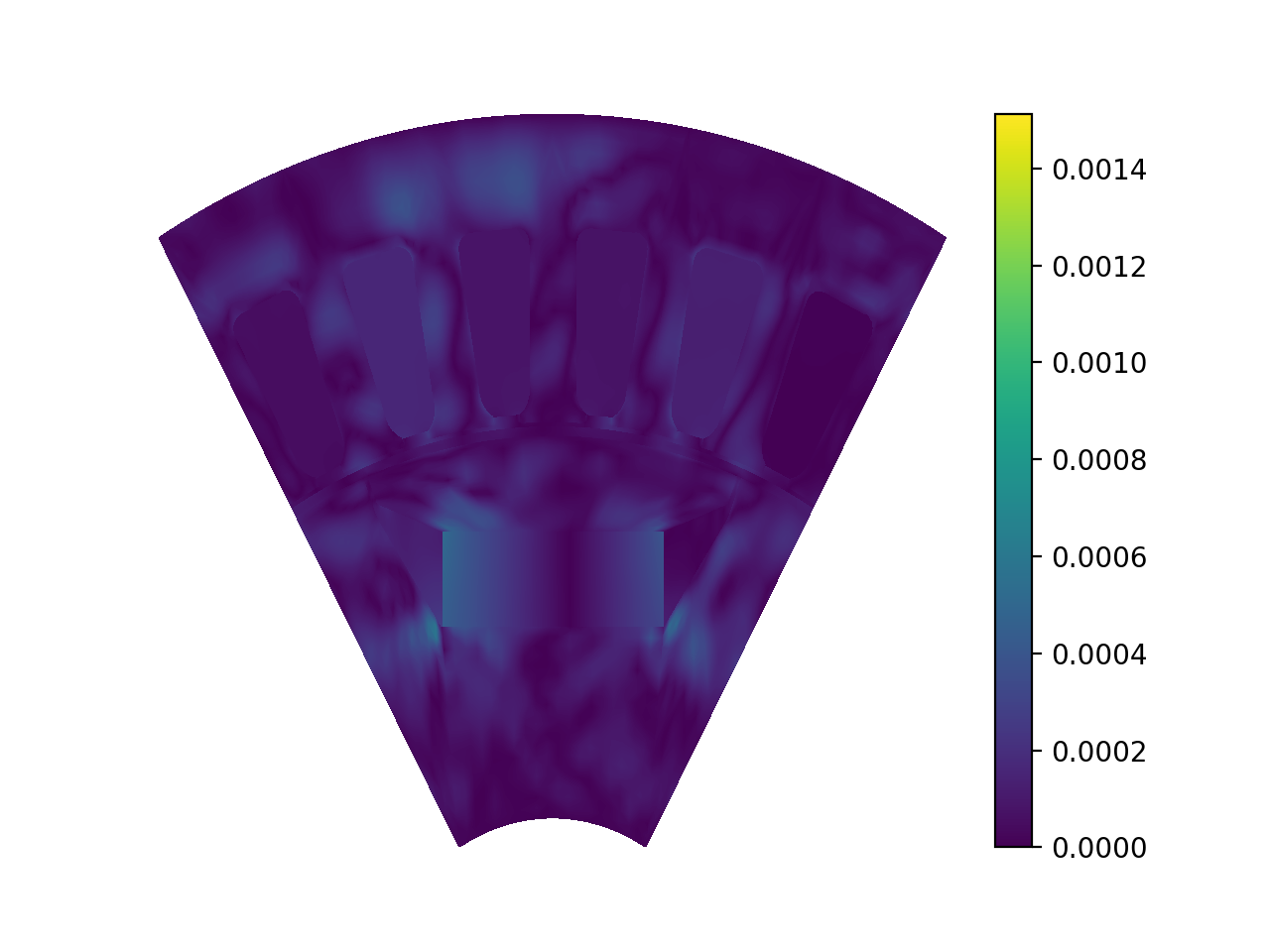}
		\caption{\gls{dg}-based neural solver.}
	\end{subfigure}%
	\begin{subfigure}[b]{.49\textwidth}
		\centering
		\includegraphics[width=1\textwidth, height=0.85\textwidth]{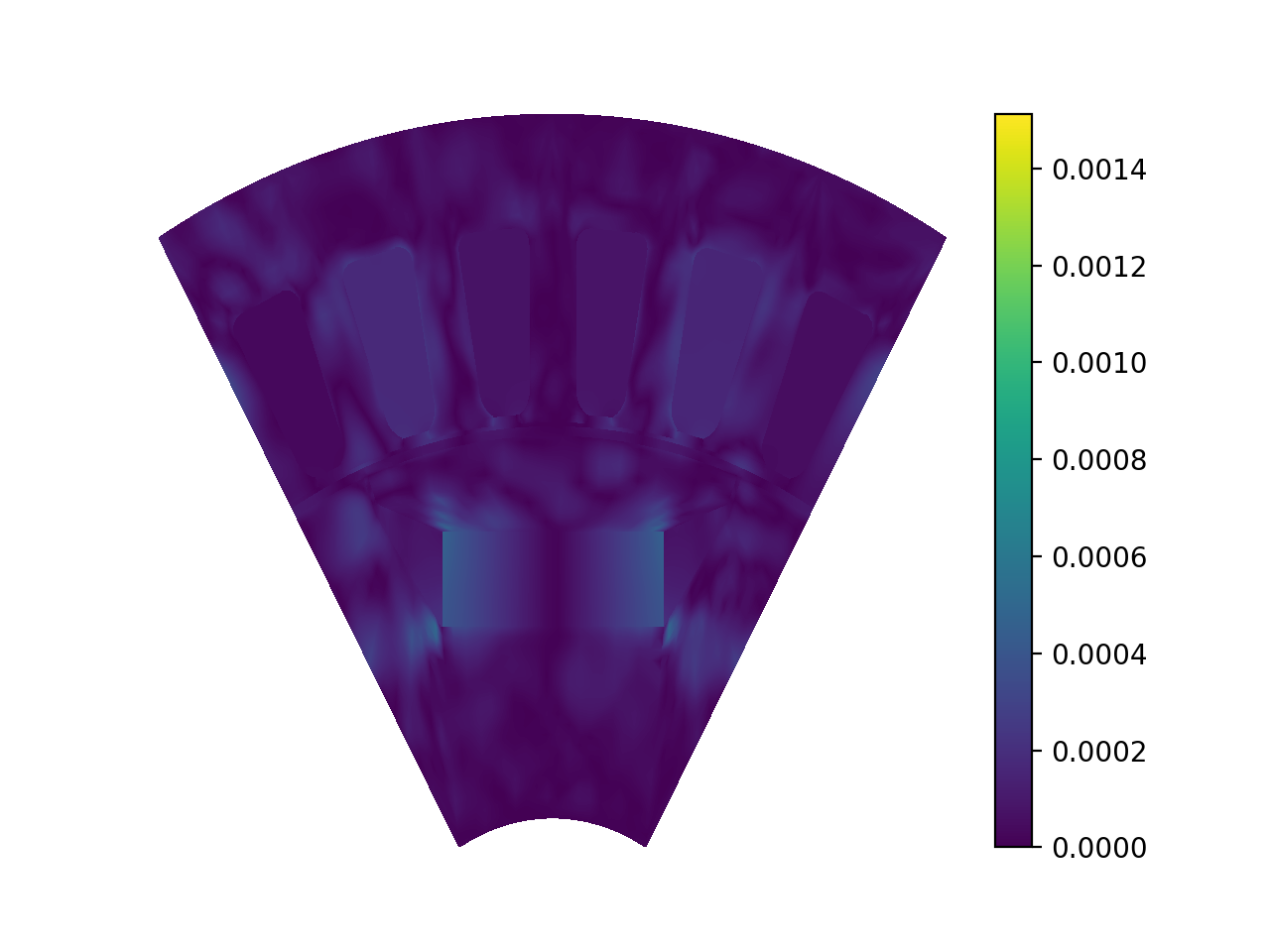}
		\caption{Coupling factors-based neural solver.}
	\end{subfigure}%
	\label{fig:IGA_ref}\\[0.5em]
	\begin{subfigure}[b]{.49\textwidth}
		\centering
		\includegraphics[width=1\textwidth, height=0.85\textwidth]{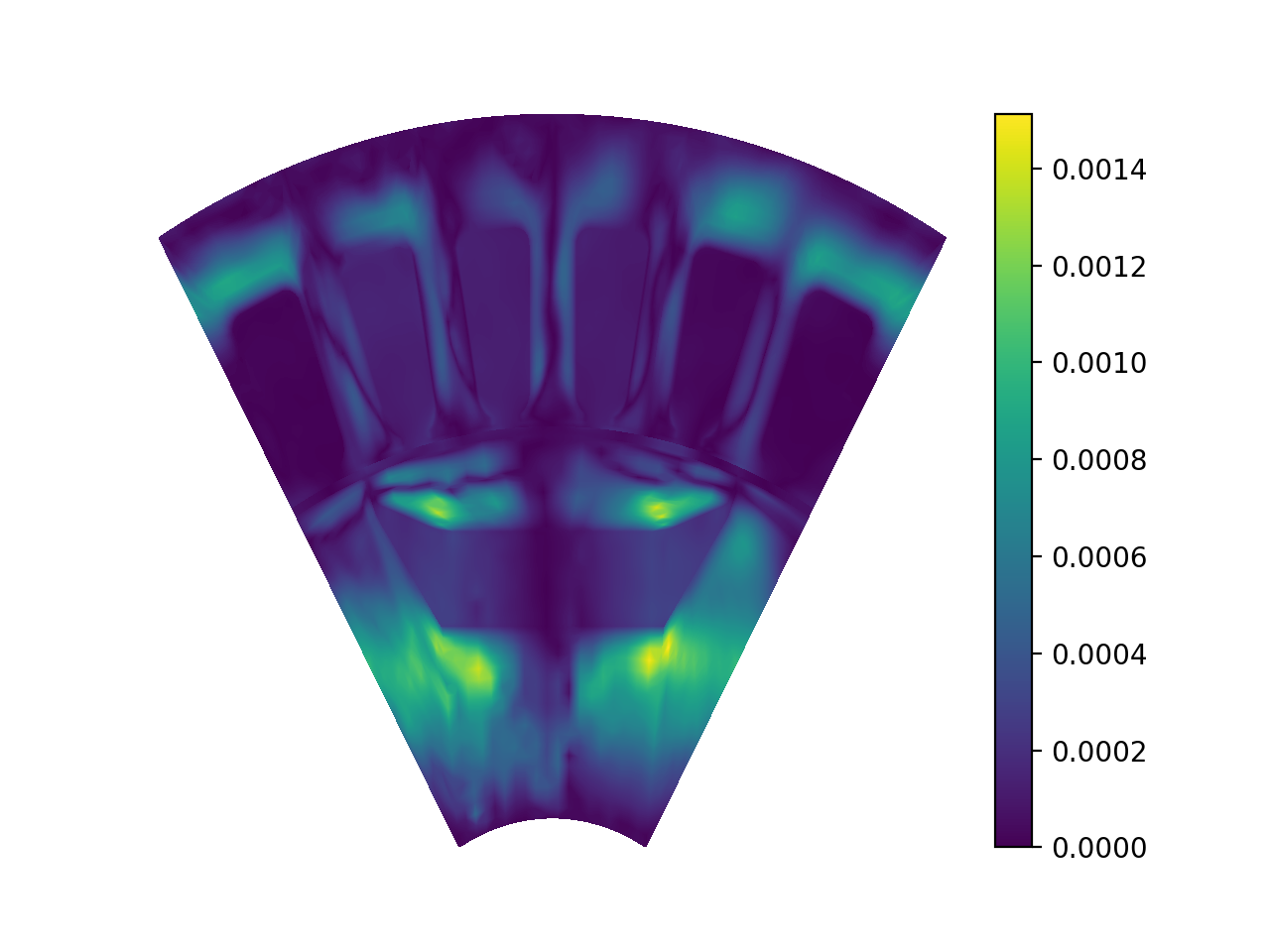}
		\caption{Single \gls{ann}-based neural solver.}
	\end{subfigure}%
	\caption{Absolute errors of the magnetic vector potential solutions provided by the \gls{dg}-based neural solver, the coupling factors-based neural solver, and the single \gls{ann}-based neural solver, with respect to the \gls{iga}-\gls{fem} reference solution.}
	\label{fig:DRM_abs_err_scope}
\end{figure}

The differences between the solutions provided by the three neural solvers become more easily distinguishable in figure~\ref{fig:DRM_abs_err_scope}, which shows the absolute error of each \gls{ann}-based solution with respect to the reference \gls{iga}-\gls{fem} over the whole domain of the \gls{pmsm}.
The absolute error is computed using $63\cdot 10^3$ randomly generated sampling points on the domain of the \gls{pmsm}, which are not included in the training data.
For the \gls{dg} and coupling factors-based neural solvers, the error remains below $5 \cdot 10^{-4}$ over the whole domain of the \gls{pmsm}, whereas the single \gls{ann}-based approximation yields worst-case errors three times higher.
In all three cases, the worst-case errors are concentrated around the corners of the permanent magnet. 
However, the single \gls{ann}-based neural solver results in comparatively high errors in other regions of the \gls{pmsm} as well, e.g. on the upper parts of the stator.

\begin{figure}[t!]
	\centering
	\begin{subfigure}[b]{.49\textwidth}
		\centering
		\includegraphics[width=1\textwidth, height=0.85\textwidth]{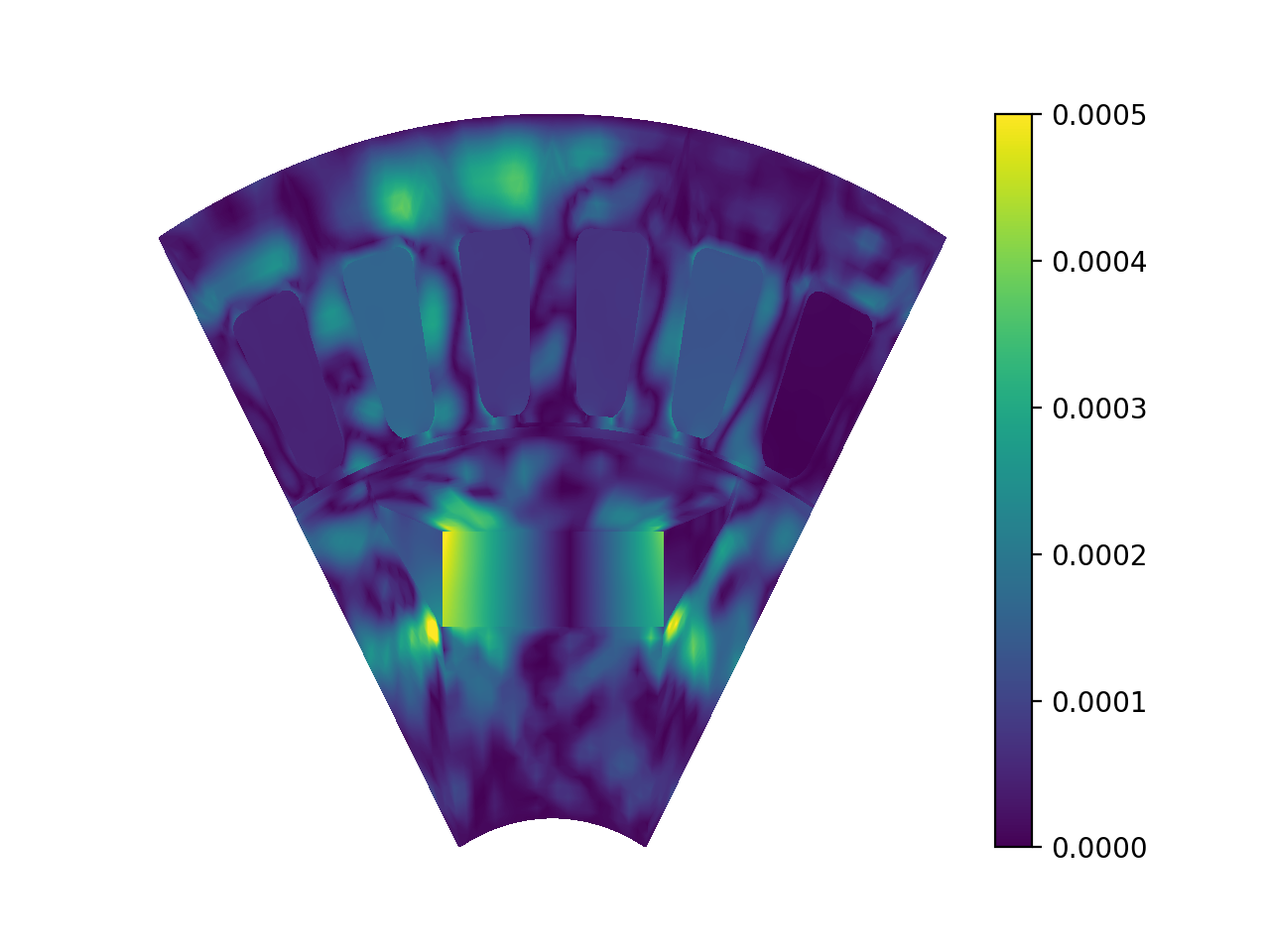}
		\caption{\gls{dg}-based neural solver.}
	\end{subfigure}%
	\begin{subfigure}[b]{.49\textwidth}
		\centering
		\includegraphics[width=1\textwidth, height=0.85\textwidth]{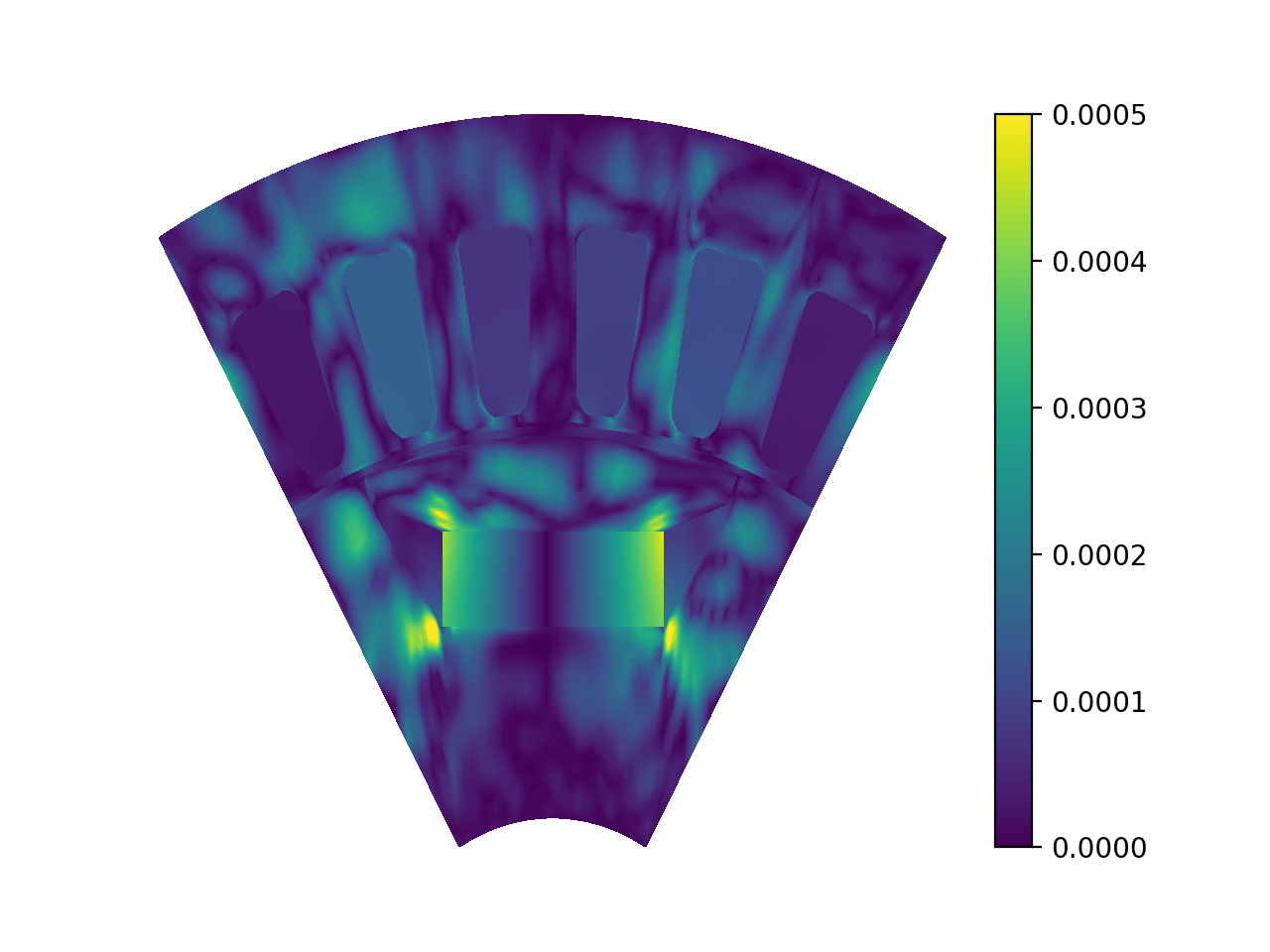}
		\caption{Coupling factors-based neural solver.}
	\end{subfigure}%
	\label{fig:IGA_ref_scope}\\[0.5em]
	\caption{Comparison between the two domain decomposition-based solvers in terms of the absolute approximation error with respect to the \gls{iga}-\gls{fem} reference solution.}
	\label{fig:DRM_abs_err}
\end{figure}

In figure~\ref{fig:DRM_abs_err}, we attempt a deeper comparison between the two domain decomposition-based neural solvers. As noted before, both neural solvers yield a maximum absolute error smaller than $5 \cdot 10^{-4}$. 
We note that the mean absolute and mean squared errors computed over the \gls{pmsm}'s domain are almost identical as well.
Nevertheless, it is evident that the error distribution over the \gls{pmsm}'s domain differs significantly. This result is similar to the one observed in section~\ref{subsec:cylinder} and can only be attributed to the different loss functions, given the fact that the \gls{ann} architectures and the hyperparameters are identical for both methods.

As a final remark for this study, we note that the reference \gls{iga}-\gls{fem} model uses non-conforming patches, thus resulting in a discontinuous magnetic vector potential along the interface $\Gamma_{\text{ag}}$ between rotor and stator. 
Contrarily, all three neural solvers yield a continuous magnetic vector potential within the air gap, which is the correct physical behavior.
Nevertheless, this discrepancy has only a minor impact on the computed errors.
Here lies a possible advantage of neural solvers over traditional solvers, which merits further investigation.

\section{Discussion, conclusion, and outlook}
\label{sec:conclusion}
In this paper, we presented a novel deep learning-based method for the solution of variational problems, which can be seamlessly integrated into \gls{cad} workflows.
An importance sampling scheme is employed for the discretization of the energy functional obtained from the variational form of the \gls{pde}, which allows to correct the loss function according to the sample distribution on the physical domain induced by \gls{cad} tools such as B-splines or \gls{nurbs}.
Complicated, multi-patch \gls{cad} domains are addressed by further modifying the loss function based on a decomposition of the computational domain and the \gls{dg} formulation.
Minimizing the \gls{dg}-based loss function allows the neural solver to produce solutions that satisfy the necessary conditions which must be met on the interfaces between subdomains and thus correctly capture sharp changes in the \gls{pde} solution.

The solver developed in this work was verified on two test cases, namely, a toy problem from \gls{em} field theory and a real-world engineering application concerning electric machine simulation. Compared to a neural solver utilizing a single \gls{ann}, the \gls{dg}-based neural solver is found to be significantly more accurate.
Importantly, the approximations obtained with the \gls{dg}-based neural solver conform to the physics of each considered test case, which is not always the case when a single \gls{ann} is used.

The \gls{dg}-based neural solver was also compared against a domain decomposition-based approach previously suggested in the literature \citep{jagtap2020conservative, jagtap2020extended}, which introduces physics-inspired coupling factors and corresponding penalty terms in the loss function. 
The two domain decomposition-based neural solvers are found to perform comparably well, however, it was observed that the different loss functions result in non-negligible differences in the distribution of the approximation error over the computational domain.
These results merit further investigations on the spatial dependency of the approximation accuracy of domain decomposition-based neural solvers, which should be extended to other approaches found in the literature as well, see e.g. \citep{kharazmi2021hp, li2019d3m, li2020deep}.
We note however that the \gls{dg} formulation, while not clearly superior to competitive approaches, offers a complete and generally applicable framework for the development of domain decomposition-based neural solvers, as follows from its extensive application in the context of the \gls{fem}. 

One drawback of the neural solver proposed in this work, as well as of neural solvers in general, concerns the duration of \gls{ann}-based simulation, which is mainly attributed to model training and typically exceeds significantly the simulation times of standard numerical solvers.
For example, the \gls{iga}-\gls{fem} solver which produced the reference solution in section~\ref{subsec:pmsm} was significantly faster than any of the employed neural solvers. 
We should note that our simulations were performed using standard \glspl{cpu}, whereas more advanced computing equipment such as \glspl{gpu} are typically necessary for achieving a good performance in \gls{ann} training.
Nevertheless, accelerated model training is necessary for neural solvers to become a truly competitive alternative to traditional numerical simulation methods. 
In connection to other persisting problems in \gls{ann}-based \gls{pde} approximation, exemplarily, the lack of convergence and stability theory, the need for hyperparameter fine-tuning, and possible overfitting problems, it should not be expected that neural solvers will replace traditional numerical solvers in the near future. This should not be surprising, as neural solvers have been revitalized only in the last few years, whereas \gls{pde} solvers based on the \gls{fem} or on other numerical schemes have been continuously developed and optimized for several decades. We note that an extensive discussion on the potential of neural solvers to replace or accelerate traditional numerical solution techniques can be found in \citep{markidis2021old}.

Nevertheless, there exist problem settings where neural solvers can be advantageous to traditional numerical methods. One relevant case is reported in \citep{berg2018unified}, where a traditional mesh-based solver fails due to the complexities of the problem's geometry, whereas the neural solver is able to provide the solution to the \gls{pde}. In the context of this work, in particular regarding the \gls{pmsm} simulation presented in section \ref{subsec:pmsm}, the numerical results revealed that the neural solvers were able to provide physics-conforming solutions in the challenging domain of the air gap between rotor and stator, where the reference \gls{iga}-\gls{fem} solver fails to do so due to the use of a non-conforming patches. Motivated by this observation, a promising research direction would be the combination of traditional and neural solvers, each applied to different regions of the computational domain where it is expected to perform better than the other.
Another possible problem setting where the utilization of neural solvers could prove to be advantageous would be for the approximation of \glspl{pde} with highly nonlinear constitutive (material) laws, which often pose significant problems for standard numerical approximation techniques \citep{galetzka2021data}. Last but not least, neural solvers are known to scale favorably for high-dimensional \glspl{pde} \citep{han2017solving, hutzenthaler2020proof}, which constitute a common bottleneck for standard numerical techniques. In the context of engineering design, typical settings where high-dimensional \glspl{pde} arise include optimization, uncertainty quantification, or parameter inference studies. It can be expected that neural solvers can mitigate the so-called curse of dimensionality and provide insights for these challenging problems.

Last, an important direction for future research concerns the development of neural solvers which preserve important physical quantities. In the context of this work, such physical quantities include continuity relations for electric and magnetic fields at material interfaces, as well as continuity of the electric potential. While these conditions are indeed found to be respected by the ANN-based solutions provided by the neural solver developed in this work, there is no a priori guarantee for that result, due to the fact that a ``soft constraints'' approach has been employed for \gls{ann} training. Such a guarantee could be achieved with the use of hard constraints or similar approach \citep{mcfall2009artificial}. In that case, one should refer to the corresponding neural solver as ``physics-constrained'' rather than ``physics-informed'', where the latter term is more appropriate for the neural solver presented in this paper. Such physics-constrained neural solvers would also compare more favorably against traditional numerical solution methods which indeed guarantee the preservation of physical quantities.

\section*{Acknowledgements}
This work was partially supported by the joint DFG/FWF Collaborative Research Centre CREATOR (CRC/TRR 361, F90) at TU Darmstadt, TU Graz, and JKU Linz. \\

\noindent Dimitrios Loukrezis and Herbert De Gersem are supported by the Graduate School Computational Engineering within the Centre for Computational Engineering at TU Darmstadt. \\

\noindent The authors would like to thank Prof. Dr. rer. nat. Sebastian Sch\"ops and Melina Merkel M.Sc. from the Chair of Computational Electromagnetics (CEM) at TU Darmstadt, as well as the former CEM members Dr.-Ing. Zeger Bontinck and Dr.-Ing. Jacopo Corno, for providing the \gls{pmsm} geometry data and the corresponding \gls{iga}-\gls{fem} solver.

\bibliographystyle{Bibliography_Style}

\bibliography{mybib.bib}{}

\appendix
\section{Geometrical and material parameters of the PMSM}
\label{sec:appendix}

\begin{figure}[t!]
\centering
\begin{tikzpicture}[scale=0.75]
   \draw (0,0) node[anchor=north]{$\cdot$};
      \draw (0,0) node[anchor=west]{$\mathbf{0}$};

   \draw [black,thick,domain=120-8.5:120] plot ({3*2.75*cos(\x)}, {3*2.75*sin(\x)});
   \draw [black,thick,domain=60:60+8.5] plot ({3*2.75*cos(\x)}, {3*2.75*sin(\x)});
   \draw [black,thick,domain=90-21:90+21] plot ({3*2.75*cos(\x)}, {3*2.75*sin(\x)});
   \draw [black,thick,domain=60:120] plot ({3*2.79375*cos(\x)}, {3*2.79375*sin(\x)});
     
   \draw [black,thick,<->, dashed, domain=120-9:60+9, >=stealth] plot ({3*2.7*cos(\x)}, {3*2.7*sin(\x)});
   \draw ({3*2.7*cos(80)}, {3*2.7*sin(80)}) node [anchor=north west]{$\delta_2$};
   
   \draw [black,thick,<->, >=stealth] ({3*2.3875*sin(75.6)*cos(90)}, {3*2.3875*sin(75.6)*sin(90)})--({3*2.75*cos(90)}, {3*2.75*sin(90)});
   \draw ({3*2.3875*sin(75.6)*cos(90)}, {3*2.3875*sin(75.6)*sin(90)+0.6}) node [anchor=east] {$d_3$};

   \draw [black,thick, dashed,<->,domain=60:60+8.5, >=stealth] plot ({3*2.7*cos(\x)}, {3*2.7*sin(\x)});
	\draw ({3*2.7*cos(60+0.5*8.5)}, {3*2.7*sin(60+0.5*8.5)}) node [anchor=north]{$\delta_1$};

   \draw [black,thick, domain=60:120] plot ({3*cos(\x)}, {3*sin(\x)});
   \draw[thick] ({3*cos(60)},{3*sin(60)}) -- ({3*2.79375*cos(60)},{3*2.79375*sin(60)}); 
   \draw[thick] ({3*cos(120)},{3*sin(120)}) -- ({3*2.79375*cos(120)},{3*2.79375*sin(120)}); 
   \draw[thick] ({3*1.9668*cos(107.57},{3*1.9668*sin(107.57)}) rectangle ({3*2.3875*cos(75.6)},{3*2.3875*sin(75.6)});
   \draw[thick] ({3*1.9668*cos(107.57)},{3*1.9668*sin(107.57)}) -- ({4*1.9668*cos(120-8.5)-0.05},{4*1.9668*sin(120-8.5)});
	\draw[thick] ({4*1.9668*cos(120-8.5)-0.05},{4*1.9668*sin(120-8.5)}) -- ({3*2.75*cos(120-8.5)}, {3*2.75*sin(120-8.5)});
	\draw[thick] ({4*1.9668*cos(120-8.5)+0.025},{4*1.9668*sin(120-8.5)}) -- ({3*2.75*cos(90+21)}, {3*2.75*sin(90+21)}); 
	\draw[thick] ({3*1.9668*cos(107.57},{3*1.9668*sin(107.57)+1.31}) -- ({4*1.9668*cos(120-8.5)+0.025},{4*1.9668*sin(120-8.5)});
	 
	\draw[thick] ({-3*1.9668*cos(107.57)},{3*1.9668*sin(107.57)}) -- ({-4*1.9668*cos(120-8.5)+0.05},{4*1.9668*sin(120-8.5)});
    \draw[thick] ({-4*1.9668*cos(120-8.5)+0.05},{4*1.9668*sin(120-8.5)}) -- ({-3*2.75*cos(120-8.5)}, {3*2.75*sin(120-8.5)});
	\draw[thick] ({-4*1.9668*cos(120-8.5)-0.025},{4*1.9668*sin(120-8.5)}) -- ({-3*2.75*cos(90+21)}, {3*2.75*sin(90+21)}); 
	\draw[thick] ({-3*1.9668*cos(107.57},{3*1.9668*sin(107.57)+1.31}) -- ({-4*1.9668*cos(120-8.5)-0.025},{4*1.9668*sin(120-8.5)});

	\draw[thick, ->, >=stealth] (0,0) -- ({3*cos(105)}, {3*sin(105)});
	\draw ({1.75*cos(105)}, {1.75*sin(105)}) node [anchor=west]{$R_{\text{rt,i}}$};
	\draw[thick, ->, >=stealth] (0,0) -- ({3*2.75*cos(118)}, {3*2.75*sin(118)});
	\draw ({3*2.4*cos(118)}, {3*2.4*sin(118)})node [anchor=west]{$R_{\text{rt,o}}$};
	\draw[thick, ->, >=stealth] (0,0) -- ({3.05*2.75*cos(58)}, {3.05*2.75*sin(58)});
	\draw ({3*2.4*cos(58)}, {3*2.4*sin(58)})node [anchor=west]{$R_{\text{st,i}}$};
	\draw[thick, ->, >=stealth] (0,0) -- ({12*cos(50)}, {12*sin(50)});
	\draw ({3*2.4*cos(50)}, {3*2.4*sin(50)})node [anchor=north west]{$R_{\text{st,o}}$};
	\draw[thick, <->, >=stealth] ({3*1.9668*cos(107.57},{3*1.9668*sin(107.57)-0.3})--({3*2.3875*cos(75.6)},{3*1.9668*sin(107.57)-0.3});
	\draw ({3*1.9668*cos(107.57)+1.9},{3*1.9668*sin(107.57)-0.3}) node [anchor=north] {$d_1$};
	\draw [thick, <->, >=stealth] ({3*2.3875*cos(75.6)-0.2},{3*1.9668*sin(107.57)})  -- ({3*2.3875*cos(75.6)-0.2},{3*2.3875*sin(75.6)});
	\draw ({3*2.3875*cos(75.6)-0.2},{3*1.9668*sin(107.57)+0.7}) node [anchor=east] {$d_2$};
	
	




	\draw [thick, ->, >=stealth] (3,0)--(3,1);
		\draw(3,0) circle (0.15 cm);

	\fill[black] (3,0) circle (0.1cm);

	\draw (3,0) node [anchor=north east]{$x_3$};
\node [anchor=west] at (3,1) {$x_2$};
\draw [thick, ->, >=stealth] (3,0)--(4,0);
\node [anchor=north] at (4,0) {$x_1$};
	
	\tikzmath{\ri=8.45; \ro=12;\del=0.1;}
	\draw[thick]({\ri*cos(120)}, {\ri*sin(120)}) to ({\ro*cos(120)}, {\ro*sin(120)});
	\draw[thick]({\ri*cos(60)}, {\ri*sin(60)}) to ({\ro*cos(60)}, {\ro*sin(60)});

	\draw[thick]({\ro*cos(120)}, {\ro*sin(120)}) to[bend left=3] ({\ro*cos(110)}, {\ro*sin(110)});
	
	 \foreach \ai in {0,10,20,30,40,50}{
	 	\draw[thick]({\ro*cos(120-\ai)}, {\ro*sin(120-\ai)}) to[bend left=3] ({\ro*cos(110-\ai)}, {\ro*sin(110-\ai)});

	\draw[thick]({\ri*cos(120-\ai)}, {\ri*sin(120-\ai)}) to[bend left=3] ({\ri*cos(120-4-\ai)}, {\ri*sin(120-4-\ai)});
	
		\draw[thick]({\ri*cos(120-6-\ai)}, {\ri*sin(120-6-\ai)}) to[bend left=3] ({\ri*cos(120-10-\ai)}, {\ri*sin(120-10-\ai)});
				
		\draw[thick]({\ri*cos(120-6-\ai)}, {\ri*sin(120-6-\ai)}) to[bend left=3] ({\ri*cos(120-10-\ai)}, {\ri*sin(120-10-\ai)});
		
				\draw[thick]({(\ri+\del)*cos(120-4-\ai)}, {(\ri+\del)*sin(120-4-\ai)}) to[bend left=3] ({(\ri+\del)*cos(120-6-\ai)}, {(\ri+\del)*sin(120-6-\ai)});

		\draw[thick]({(\ri+14*\del)*cos(120-2.75-\ai)}, {(\ri+14*\del)*sin(120-2.75-\ai)}) to[bend left=3] ({(\ri+14*\del)*cos(120-7.25-\ai)}, {(\ri+14*\del)*sin(120-7.25-\ai)});
		\draw[thick]({(\ri+23*\del)*cos(120-3.5-\ai)}, {(\ri+23*\del)*sin(120-3.5-\ai)}) to[bend left=3] ({(\ri+23*\del)*cos(120-6.5-\ai)}, {(\ri+23*\del)*sin(120-6.5-\ai)});
		
		
		\draw[thick]({(\ri+\del)*cos(120-4-\ai)}, {(\ri+\del)*sin(120-4-\ai)}) to[bend left=20] ({(\ri+2*\del)*cos(120-3-\ai)}, {(\ri+2*\del)*sin(120-3-\ai)});

		\draw[thick]({(\ri+2*\del)*cos(120-3-\ai)}, {(\ri+2*\del)*sin(120-3-\ai)}) to ({(\ri+22*\del)*cos(120-2.5-\ai)}, {(\ri+22*\del)*sin(120-2.5-\ai)});

		\draw[thick]({(\ri+22*\del)*cos(120-2.5-\ai)}, {(\ri+22*\del)*sin(120-2.5-\ai)}) to[bend left=30] ({(\ri+23*\del)*cos(120-3.5-\ai)}, {(\ri+23*\del)*sin(120-3.5-\ai)});

		
		\draw[thick]({(\ri+\del)*cos(120-6-\ai)}, {(\ri+\del)*sin(120-6-\ai)}) to[bend right=20] ({(\ri+2*\del)*cos(120-7-\ai)}, {(\ri+2*\del)*sin(120-7-\ai)});

		\draw[thick]({(\ri+2*\del)*cos(120-7-\ai)}, {(\ri+2*\del)*sin(120-7-\ai)}) to ({(\ri+22*\del)*cos(120-7.5-\ai)}, {(\ri+22*\del)*sin(120-7.5-\ai)});

		\draw[thick]({(\ri+22*\del)*cos(120-7.5-\ai)}, {(\ri+22*\del)*sin(120-7.5-\ai)}) to[bend right=30] ({(\ri+23*\del)*cos(120-6.5-\ai)}, {(\ri+23*\del)*sin(120-6.5-\ai)});

	\draw[thick]({\ri*cos(120-4-\ai)}, {\ri*sin(120-4-\ai)}) to[bend left=3] ({(\ri+\del)*cos(120-4-\ai)}, {(\ri+\del)*sin(120-4-\ai)});
		
		\draw[thick]({\ri*cos(120-6-\ai)}, {\ri*sin(120-6-\ai)}) to[bend left=3] ({(\ri+\del)*cos(120-6-\ai)}, {(\ri+\del)*sin(120-6-\ai)});

	\draw [black,thick,<->, >=stealth] ({(\ri+\del)*cos(82)}, {(\ri+\del)*sin(82)})--({(\ri+14*\del)*cos(81.5)}, {(\ri+14*\del)*sin(81.5)});
   \draw ({(\ri+7*\del)*cos(82)}, {(\ri+7*\del)*sin(82)}) node [anchor=west] {$l_2$};

	\draw [thick,->, >=stealth] ({(\ri+4*\del)*cos(75)}, {(\ri+4*\del)*sin(75)})--({(\ri+\del)*cos(75)}, {(\ri+\del)*sin(75)});

	\draw [thick,->, >=stealth] ({(\ri+-3*\del)*cos(75)}, {(\ri+-3*\del)*sin(75)})--({8.45*cos(75)}, {8.45*sin(75)});

\draw ({(\ri+6*\del)*cos(77)}, {(\ri+6*\del)*sin(77)}) node [anchor=west] {$l_1$};

	\draw [black,thick,<->, >=stealth] ({(\ri+14*\del)*cos(81.5)}, {(\ri+14*\del)*sin(81.5)})--({(\ri+23*\del)*cos(81.25)}, {(\ri+23*\del)*sin(81.25)});
	   \draw ({(\ri+18*\del)*cos(81.7)}, {(\ri+18*\del)*sin(81.7)}) node [anchor=west] {$l_3$};

\draw [black,thick,<->, >=stealth] ({(\ri+23*\del)*cos(85)}, {(\ri+23*\del)*sin(85)})--({(12)*cos(85)}, {(12)*sin(85)});
   \draw ({(11.3)*cos(85.25)}, {(11.3)*sin(85.25)}) node [anchor=west] {$l_4$};

		\draw[thick, <->, >=stealth, dashed]({(\ri+15*\del)*cos(120-2.75-40)}, {(\ri+15*\del)*sin(120-2.75-40)}) to[bend left=3] ({(\ri+15*\del)*cos(120-7.25-40)}, {(\ri+15*\del)*sin(120-7.25-40)});
\draw ({(\ri+16*\del)*cos(120-3.5-40)}, {(\ri+16*\del)*sin(120-3.5-40)}) node [anchor=west] {$\delta_4$};

\draw[thick, <->, >=stealth, dashed]({(\ri+22*\del)*cos(120-7.5-40)}, {(\ri+22*\del)*sin(120-7.5-40)}) to[bend left=3] ({(\ri+22*\del)*cos(120-7.25-45.25)}, {(\ri+22*\del)*sin(120-7.25-45.25)});
\draw ({(\ri+23*\del)*cos(120-3.5-45.25)}, {(\ri+23*\del)*sin(120-3.5-45.25)}) node [anchor=west] {$\delta_5$};

\draw[thick, <->, >=stealth, dashed]({(\ri+0.5*\del)*cos(120-6-40)}, {(\ri+0.5*\del)*sin(120-6-40)}) to[bend left=3] ({(\ri+0.5*\del)*cos(120-8.75-45.25)}, {(\ri+0.5*\del)*sin(120-8.75-45.25)});
\draw ({(\ri+1.5*\del)*cos(120-3-45.25)}, {(\ri+1.5*\del)*sin(120-3-45.25)}) node [anchor=west] {$\delta_3$};




	};

\draw[thick] ({8*cos(120)},{8*sin(120)})--({9*cos(120)},{9*sin(120)});
	\draw[thick] ({8*cos(60)},{8*sin(60)})--({9*cos(60)},{9*sin(60)});
	
	\draw ({9*cos(130)}, {9*sin(130)}) node [anchor=center]{$\Gamma_{\text{ag}}$};
	\draw[very thick, ->, >=stealth]({9*cos(130)+0.4}, {9*sin(130)}) to[bend left=30] ({9*cos(130)+2}, {9*sin(130)+0.6});

\end{tikzpicture}
    \caption{One-sixth of the \gls{pmsm}'s geometry, along with its geometric parameters.}
    \label{fig:motor_seg_geo}
\end{figure}
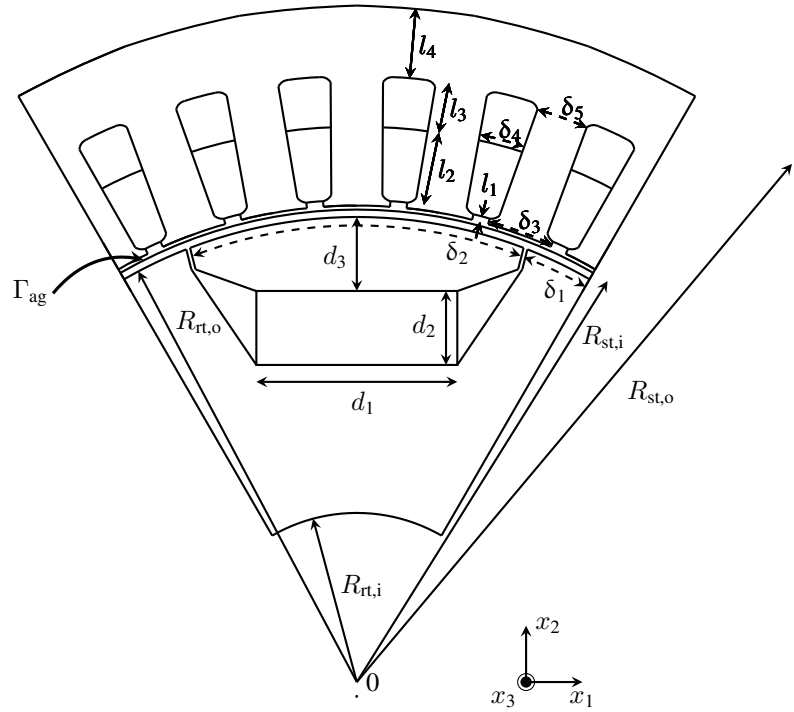

\begin{table}[b!]
	\centering
	\begin{tabular}{l c c c c}
		\hline\hline
		Description & Symbol & Value & Units \\ [0.5ex] 
		\hline
		Inner rotor radius &$R_{\text{rt,i}} $&$16$ &\si{mm} \\
		Outer rotor radius&$R_{\text{rt,o}}$&$44$ &\si{mm} \\
		Radius of $\Gamma_{\text{ag}}$& $R_{\text{ag}}$ & $44.7$ &\si{mm} \\
		Magnet width &$d_1$& $19$ &\si{mm} \\
		Magnet height &$d_2$& $7$ &\si{mm} \\
		Depth of magnet in rotor& $d_3$ & $7$ &\si{mm} \\
		Inner stator radius &$R_{\text{st,i}} $&$45$ &\si{mm} \\
		Outer stator radius&$R_{\text{st,o}}$&$67.5$ &\si{mm} \\
		& $\delta_1$ & $8.5$ &\si{^\circ} \\
		& $\delta_2$ & $42$ &\si{^\circ} \\
		& $\delta_3$ & $7$ &\si{^\circ} \\
		& $\delta_4$ & $5.7$ &\si{^\circ} \\
		& $\delta_5$ & $4$ &\si{^\circ} \\
		& $l_1$ & $0.6$ &\si{mm} \\
		& $l_2$ & $5.4$ &\si{mm} \\
		& $l_3$ & $5$ &\si{mm} \\
		& $l_4$ & $8.2$ &\si{mm} \\
		\hline
	\end{tabular}
	\caption{Geometrical parameters of the PMSM, as depicted in figure~\ref{fig:motor_seg_geo}.}
	\label{table:rotor_dim}
\end{table}

\begin{table}[b!]
	\centering
	\begin{tabular}{l c c c c}
		\hline\hline
		Description & Symbol & Value & Units \\ [0.5ex] 
		\hline
		Relative reluctivity of iron & $\nu_{\text{Fe}}$  &$1/500$ & -- \\
		Relative reluctivity of copper & $\nu_{\text{Cu}}$  &$1$ & -- \\
		Relative reluctivity of PM & $\nu_{\text{PM}}$  &$1/1.05$ & -- \\
		Remanent magnetic field of PM & $B_\text{r}$  &$0.94$ &\si{T} \\[1ex]
		\hline
	\end{tabular}
	\caption{Material parameters of the PMSM.}
	\label{table:mat_par}
\end{table}

\end{document}